\documentclass[twocolumn]{aastex62}

\usepackage{subfigure}
\usepackage{url}
\usepackage{hyperref}
\usepackage{longtable}
\usepackage{natbib}
\usepackage{amsmath}
\usepackage{listings}
\usepackage[normalem]{ulem}
\usepackage{bm}
\usepackage{comment}
\usepackage{soul}
\usepackage{array}
\newcolumntype{P}[1]{>{\centering\arraybackslash}p{#1}}
\newcolumntype{M}[1]{>{\centering\arraybackslash}m{#1}}

\usepackage{xcolor, fontawesome}
\definecolor{twitterblue}{RGB}{64,153,255}

\newcommand{\twitter}[1]{\href{https://twitter.com/#1 }{\textcolor{twitterblue}{\faTwitter}\,\tt \textcolor{twitterblue}{@#1}}}

\newcommand{\github}[1]{\href{https://github.com/#1 }{\textcolor{black}{\faGithub}\,\tt \textcolor{black}{@#1}}}

\definecolor{Code}{rgb}{0,0,0}
\definecolor{Decorators}{rgb}{0.5,0.5,0.5}
\definecolor{Numbers}{rgb}{0.5,0,0}
\definecolor{MatchingBrackets}{rgb}{0.25,0.5,0.5}
\definecolor{Keywords}{rgb}{1,0,0}
\definecolor{self}{rgb}{0,0,0}
\definecolor{Strings}{rgb}{0,0.63,0}
\definecolor{Comments}{rgb}{0,0.63,1}
\definecolor{Backquotes}{rgb}{0,0,0}
\definecolor{Classname}{rgb}{0,0,0}
\definecolor{FunctionName}{rgb}{0,0,0}
\definecolor{Operators}{rgb}{0,0,0}
\definecolor{Background}{rgb}{0.98,0.98,0.98}
\definecolor{Booleans}{rgb}{0.572,0,0.572}
\definecolor{BuiltinFunction}{rgb}{0.572,0,0.572}
\definecolor{BuiltinConstant}{rgb}{0.572,0,0.572}
\definecolor{Asterisk}{rgb}{0.670,0,1}

\lstdefinelanguage{Python}{
    	numbers=left,
    	numberstyle=\footnotesize,
    	numbersep=7pt,
    	xleftmargin=1.26em,
    	framextopmargin=2em,
    	framexbottommargin=2em,
    	showspaces=false,
    	showtabs=false,
    	showstringspaces=false,
    	frame=l,
    	tabsize=4,
    	stepnumber=1,
	basicstyle=\small\ttfamily,
    	backgroundcolor=\color{Background},
	stringstyle=\ttfamily\color{Strings},
	morekeywords={import,from,class,def,while,if,in,elif,else,not,or,print,break,continue,return,access,as,except,exec,finally,global,import,lambda,pass,print,raise,try,assert},
    	keywordstyle={\color{Keywords}\bfseries}, 
	otherkeywords={[2]*},
	keywordstyle={[2]\color{Asterisk}},
}

\usepackage{color}

\newcommand{\shrug}{\texttt{\raisebox{0.75em}{\char`\_}\char`\\\char`\_\kern-0.5ex(\kern-0.25ex\raisebox{0.25ex}{\rotatebox{45}{\raisebox{-.75ex}"\kern-1.5ex\rotatebox{-90})}}\kern-0.5ex)\kern-0.5ex\char`\_/\raisebox{0.75em}{\char`\_}}}

\newcommand{\ktwo}{{\it K2}}
\newcommand{\tess}{{\it TESS}}

\newcommand{\Gaia}{{\it Gaia}}

\newcommand{\vsini}{{$v \sin i$}}
\newcommand{\teff}{$T_{\rm{eff}}$}

\newcommand{\stella}{\texttt{stella}}
\newcommand{\banyan}{\texttt{BANYAN-$\Sigma$}}

\newcommand{\logg}{{log(g)}}

\newcommand{\chicago}{Department of Astronomy and Astrophysics, University of
Chicago, 5640 S. Ellis Ave, Chicago, IL 60637, USA}

\newcommand{\nsf}{NSF Graduate Research Fellow}

\newcommand{\unsw}{School of Physics, University of New South Wales, Sydney, NSW 2052, Australia}
\newcommand{\udash}{UNSW Data Science Hub, University of New South Wales, Sydney, NSW 2052, Australia}

\newcommand{\austin}{Department of Astronomy, The University of Texas at Austin, 2515 Speedway Boulevard, Austin, TX 78712, USA}

\newcommand{\flatiron}{Center for Computational Astrophysics, Flatiron Institute, 162 Fifth Ave, New York, NY 10010, USA}

\newcommand{\lco}{Las Cumbres Observatory, 6740 Cortona Drive, Suite 102, Goleta, CA 93117, USA}

\newcommand{\tokyo}{Department of Astronomy, University of Tokyo, 7-3-1 Hongo, Bunkyo-ku, Tokyo 113-0033, Japan}

\newcommand{\amnh}{Department of Astrophysics, American Museum of Natural History, New York, NY 10024, USA}

\newcommand{\vtau}{V1298\,Tau}
\newcommand{\halpha}{H$\alpha$}

\newcommand{\singly}{\textsc{i}}
\newcommand{\doubly}{\textsc{ii}}
\newcommand{\starry}{\texttt{starry}}
\newcommand{\calcium}{Ca\,\textsc{ii}}
\newcommand{\calciumhk}{Ca\,\textsc{ii}\,H~\&~K}
\newcommand{\helium}{He\,\textsc{i}}
\newcommand{\sodium}{Na\,\textsc{i}}
\newcommand{\lithium}{Li\,\textsc{i}}
\newcommand{\magnesium}{Mg\,\textsc{i}}
\newcommand{\target}{\vtau\,c}
\newcommand{\angstrom}{\mbox{\normalfont\AA}}
\newcommand{\obliquity}{$5^\circ \pm 15^\circ$}

\definecolor{linkcolor}{rgb}{0.41960784, 0.40784314, 0.39607843}
\newcommand{\codeicon}{{\color{linkcolor}\faCloudDownload}}

\newcommand{\reductcloud}{\href{https://github.com/afeinstein20/doppler_tomography/blob/paper-version/demo_notebooks/reduction_steps.ipynb}{\codeicon}}
\newcommand{\ewcloud}{\href{https://github.com/afeinstein20/doppler_tomography/blob/paper-version/demo_notebooks/means.ipynb}{\codeicon}}
\newcommand{\mcmccloud}{\href{https://github.com/afeinstein20/doppler_tomography/blob/paper-version/dt_code/horus_mcmc.py}{\codeicon}}
\newcommand{\spotcloud}{\href{https://github.com/afeinstein20/doppler_tomography/blob/paper-version/notebooks/spot_model.ipynb}{\codeicon}}
\newcommand{\clvcloud}{\href{https://github.com/afeinstein20/doppler_tomography/blob/paper-version/notebooks/clv.ipynb}{\codeicon}}

\newcommand{\davidcite}{David et al. (in prep)}
\newcommand{\davidcitep}{(David et al. in prep)}

\newcommand{\johnsoncitep}{(Johnson et al. in prep)}

\newcommand{\livingstoncitep}{(Livingston priv. comm.)}

\begin{document}
\title{\halpha\ and \calcium\ Infrared Triplet Variations During a Transit of the 23 Myr Planet \vtau\,c}

\shorttitle{\vtau\ Stellar Activity} 
\shortauthors{Feinstein et al.}

\author[0000-0002-9464-8101]{Adina~D.~Feinstein}
\altaffiliation{\nsf}
\affiliation{\chicago}

\author[0000-0001-7516-8308]{Benjamin~T.~Montet}
\affiliation{\unsw}
\affiliation{\udash}

\author[0000-0002-5099-8185]{Marshall~C.~Johnson}
\affiliation{\lco}

\author[0000-0003-4733-6532]{Jacob~L.~Bean}
\affiliation{\chicago}

\author[0000-0001-6534-6246]{Trevor~J.~David}
\affiliation{\flatiron}
\affiliation{\amnh}

\author[0000-0002-4020-3457]{Michael~A.~Gully-Santiago}
\affiliation{\austin}

\author[0000-0002-4881-3620]{John~H.~Livingston}
\affiliation{\tokyo}

\author[0000-0002-0296-3826]{Rodrigo Luger}
\affiliation{\flatiron}

\correspondingauthor{Adina~D.~Feinstein;\\ \twitter{afeinstein20}; \github{afeinstein20};} \email{afeinstein@uchicago.edu}


\features{Affiliated Python scripts/Jupyter notebooks \href{https://github.com/afeinstein20/doppler_tomography/tree/paper-version}{\codeicon}}

\begin{abstract}

Young transiting exoplanets ($<$~100~Myr) provide crucial insight into atmospheric evolution via photoevaporation. However, transmission spectroscopy measurements to determine atmospheric composition and mass loss are challenging due to the activity and prominent stellar disk inhomogeneities present on young stars. We observed a full transit of \target, a 23~Myr, 5.59~R\textsubscript{$\oplus$} planet orbiting a young K0-K1.5 solar analogue with GRACES on Gemini-North. We were able to measure the Doppler tomographic signal of \target\ using the \calcium\ infrared triplet (IRT) and find a projected obliquity of $\lambda = $~\obliquity. The tomographic signal is only seen in the chromospherically driven core of the \calcium~IRT, which may be the result of star-planet interactions.  Additionally, we find that excess absorption of the \halpha\ line decreases smoothly during the transit. While this could be a tentative detection of hot gas escaping the planet, we find this variation is consistent with similar timescale observations of other young stars that lack transiting planets over similar timescales. We show this variation can also be explained by the presence of starspots with surrounding facular regions. More observations both in- and out-of the transits of \target\ are required to determine the nature of the \calcium~IRT and \halpha\ line variations.

\end{abstract}

\section{Introduction}\label{sec:intro}

Young planets provide valuable insights into the early stages of planet formation and evolution. Through transit missions like \ktwo\ \citep{Howell14} and \tess\ \citep[Transiting Exoplanet Survey Satellite;][]{Ricker14} the population of young transiting planets (here defined as $<$~100~Myr) has grown to approximately a dozen \citep{david16, mann16, benatti19, newton19, mann20, plavchan20, rizzuto20}. These planets are key to understanding planetary migration, atmospheric evolution, and measuring atmospheric mass-loss rate due to photoevaporation.

\subsection{Obliquity Measurements}

Close-in transiting exoplanets have brought into question the mechanisms in which they get there. Measuring the obliquity, $\lambda$, or the relative angle between the spin of the star and orbit of the planet \citep{Lai14}, can yield insights into different migratory paths. A low-obliquity hints at a smooth disk migration history \citep{goldreich79, ford14}, while binary companions or nearby stars in stellar birth clusters can torque planets to a high obliquity over millions of years \citep{Fabrycky07}. 

Such measurements have only just begun for young systems. The recently discovered 20~Myr planet AU\,Mic\,b \citep{plavchan20} has been the subject of significant follow-up. All independent studies have measured a low obliquity \citep{addison20, hirano20, martioli20, palle20}, suggesting the planet formed beyond the ice-line within the protoplanetary disk and experienced a smooth migration inward. 

Similar studies of spin-orbit alignment have been completed for DS\,Tuc\,Ab, a planet in a slightly older (35-45~Myr) system, in a known binary \citep{newton19, montet20, zhou20}. The results from both studies concluded the planet has a low obliquity and that the migration was not significantly torqued through Kozai-Lidov oscillations by DS Tuc B. This is additionally supported by the fact that the age of the system is younger than the timescale required for Kozai-Lidov interactions \citep{montet20}. \vtau\ has no known companions and thus any spin-orbit misalignment would hint towards potential torquing of the disk due to neighboring stars in the birth cluster.

\subsection{Photoevaporation \& Young Stellar Activity}

Photoevaporation of young close-in planets is one of the leading explanations for the ``radius gap" of transiting planets \citep{owen17, fulton17, vaneylen18, Bean21}. This gap is between a bimodal distribution of small planets, centered at 1.3 and 2.4 $R_\oplus$ with a distinct lack of planets in between. However, early studies did not account for the ages of the systems; \cite{berger20} presented the first analysis of the planet-radius distribution while accounting for such ages. They found tentative evidence of an age dependence of the planet radius distribution, demonstrating that the number of super-Earths (R/R$_\oplus <1.8$) increases with age, assuming a fixed radius gap location. This could be attributed to a core-powered mass-loss mechanism \citep{gupta19}, which may continue for the first 1~Gyr \citep{rogers21}. However, the general contributions of atmospheric removal from photoevaporation and core-powered mass-loss is currently unknown. While photoevaporation rate is a function of flux, these young transiting planets have short orbital periods and allow us to probe the most dramatic environments for atmospheric removal.

Atmospheric photoevaporation is driven by the high energy irradiation from young stars \citep{lammer03, owen19}. The effects are particularly dramatic in the first 100~Myr where stellar XUV emission is higher and planets are still contracting, which makes their atmospheres more vulnerable to mass loss \citep{Preibisch2005, feigelson07}. Using \textit{GALEX} observations, \cite{Shkolnik14} found that early M stars remain at high, saturated levels of NUV and FUV flux for the first $\sim$650~Myr. The same relation was found of K stars \citep{richey-yowell19}.
The activity lifetime for low-mass stars is longer than that of solar-type stars. However, the higher levels of XUV flux makes photoevaporation more efficient for planets around low-mass stars \citep{rogers21}, thus the evolution of planetary radii is faster \citep{owen13}. 

An increase in magnetic activity leads to elevated UV/X-Ray radiation in young stars, manifests itself in larger spot coverage, and results in increased photometric variability \citep{feigelson99}. Several detailed studies of T Tauri stars have estimated that spots cover 29-41\% of the less spotted hemisphere and 61-67\% of the more spotted hemisphere \citep{grankin99}, with the extreme case of LkCa 4 with of $\sim 80$\% spot coverage \citep{gully17}. Larger studies of young stars have also seen a lack of phase dependence for flares, another proxy for magnetic activity, indicating similar levels of high spot coverage across both hemispheres \citep{feinstein20}. For detailed spectroscopic studies of young planets (e.g. studying their spin-orbit alignment or atmospheric characterization), it is crucial to understand the underlying starspot and active region coverage.

\subsection{\vtau}

\vtau\ is a 23~Myr pre-main sequence K star (M$_\star$/M$_\odot$ = 1.101; R$_\star$/R$_\odot$ = 1.345) hosting four known transiting planets \citep{david19_v1298all, david19_v1298b} observed in \ktwo\ Campaign 4 \citep{Howell14}. \vtau\ is the first known young transiting multi-planet system. \vtau\ was first identified as a young star via strong X-ray emission observed with \textit{ROSAT} and strong \lithium~670.8~nm absorption \citep{wichmann96, wichmann00}. \vtau\ belongs to a young association in the foreground of the Taurus-Auriga star-forming complex \citep{oh17} with an estimated age of 20-30 Myr based on comparison to empirical and theoretical isochrones \citep{luhman18, david19_v1298b}. \target\ ($R\sim5.59R_\oplus$) and d ($R\sim6.41R_\oplus$) are predicted to evolve to sub-Neptunes ($< 4 R_\oplus$) or super-Earths ($< 2 R_\oplus$), depending on the XUV evolution of the host star, after 5~Gyr \citep{david19_v1298all, poppenhaeger20}. The light curve for \vtau\ also shows strong starspot variability, with a peak-to-trough amplitude of $\sim 6$\% \citep{david19_v1298b}. Therefore, any in-transit spectra need to be carefully searched for the signatures of contaminating ARs and spots.

\vtau\,b is seen to be well aligned through both a clear Rossiter-McLaughlin signal and Doppler tomographic analyses \johnsoncitep. Additionally, \davidcite\ reports detections of excess \halpha\ absorption during transits of planets \vtau\,b, hinting at highly extended atmospheres, which is expected for young planets.

Particularly, the excess absorption dominates in the red wing of the line and enhanced absorption at transit ingress of \vtau\,b, which is confined to the core of the line. As the transit continues, there is an excess in absorption in the wings of the line and more so on the red side. The excess absorption during ingress could be the planet occulting a bright stellar surface or limb feature. Several observations during transits of \target\ and \vtau\,d suggest tentative evidence for excess \halpha\ absorption \davidcitep. Observations do not span full transits, and therefore it is challenging to attribute the excess to the planets alone.

Here, we present a spectroscopic analysis during a full transit of \target, the innermost known planet in the system. We measure the spin-orbit alignment of the planet via Doppler Tomography and explore potential signs of atmospheric escape through changes in \halpha. The paper is presented as follows: Section~\ref{sec:obs} describes our observations and data analysis through both Doppler tomography and detailed line work. We present our results in Section~\ref{sec:results}. Section~\ref{sec:discussion} discusses potential sources of the \halpha\ variations observed including center-to-limb variations, underlying starspots and faculae, and an extended planetary atmosphere from \target, and the source of the \calcium\ Infrared Triplet (IRT). We present our interpretation and future work in Section~\ref{sec:future}. We conclude by putting \target\ in context with other young planets and a summary of our key findings in Section~\ref{sec:conclusion}.

\section{Observations \& Methodology}\label{sec:obs}

Due to the youth and high levels of stellar activity for \vtau, we took several approaches to interpreting our observations. We observed other young active stars to compare spectral line variations to those seen in \vtau. An updated ephemeris for \target\ was obtained from recent \textit{Spitzer} observations \livingstoncitep.

\subsection{GRACES Observations}

We obtained data with GRACES \citep[Gemini Remote Access to CFHT ESPaDOnS Spectrograph;][]{chene14} mounted on Gemini North at the Gemini Observatory on UT 2020 January 22 through the Fast Turnaround program. GRACES has a wavelength coverage of 400-1000~nm and a resolving power of R $\sim$ 81,000. To obtain an optimal signal-to-noise of approximately 300 at 600~nm, we observed \vtau\ at 360~second exposures. \vtau\ was monitored for approximately 6~hours in total, resulting in 53 exposures for this analysis. Observations were taken under photometric conditions. 
Our observations cover the entire transit of \target. Five exposures at the beginning and six at the end are out-of-transit (OOT). We note that a transit of \vtau\,d ended approximately 12~minutes before our observations began. 

\subsubsection{Data Reduction \reductcloud}\label{subsec:reduction}

Raw and reduced spectra were provided by Gemini-N. The spectra were reduced using the OPERA\footnote{Open source Pipeline for ESPaDOnS Reduction and Analysis} pipeline \citep{martioli12, teeple14}. The OPERA pipeline completes several pre-reduction stages including creating master calibration images, a bad-pixel mask, and calibrating the instrument profile, aperture, and wavelength. The pipeline then extracts spectra from the original CCD images, and normalizes and telluric-corrects the spectra. However, we found a custom reduction of the data provides cleaner spectra for our analysis, via visual inspection. 

We extracted each spectrum through the box-extraction method. We dark-subtracted and flat-corrected the spectra using master frames created by taking a median of all available dark and flat exposures. Cosmic rays were identified and removed using \texttt{ccdproc} \citep{ccdproc}, an extension of \texttt{astropy} \citep{astropy:2013, astropy18} that tracks bad pixels in images. 

To extract individual orders, we used five equally spaced columns across the frame and identified local minima across each column. The minima corresponded to regions between the orders. A second-degree polynomial was then fit to each of the five points in a given row to trace and extract the order. We visually inspected that each model corresponded to a single order, and did not cross over into a different order. We discretized the models and ensured the box was the same width across the orders. We then box-extracted 23 orders in each exposure.

\begin{figure}[!ht]
\begin{center}
\includegraphics[width=0.46\textwidth,trim={0.25cm 0 0 0}]{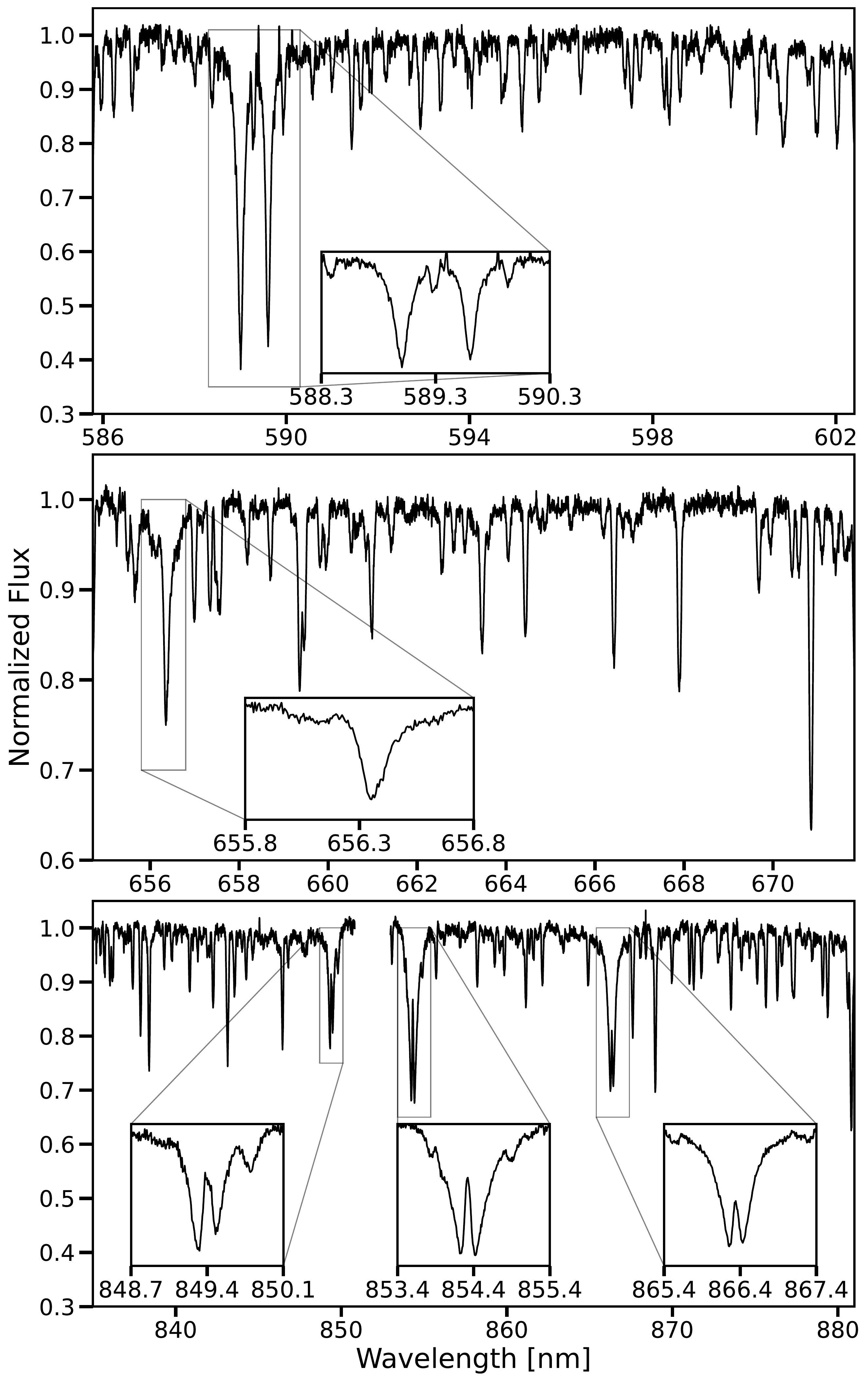}
\caption{Example extracted spectra for \vtau\ obtained with GRACES on Gemini-North. The bottom panel contains spectra from two different orders to highlight the \calcium~IRT at 849.8, 854.2, and 866.2~nm. Inset panels highlight relevant lines used in this analysis including the \sodium\ doublet (top), \halpha\ (middle), and \calcium~IRT (bottom). \label{fig:spectra}}
\end{center}
\end{figure}

To correct for the blaze, we removed the first and last 350 data points and created a 95\textsuperscript{th}-percentile filter using \texttt{scipy} with a window length of 150 data points. For stars with broad lines, such as \vtau, this is a standard procedure for identifying the blaze when there is no distinct continuum available \citep{pineda13, montet15a}. An 8\textsuperscript{th}-degree polynomial was fit to this filter and divided out of the spectra, resulting in our final corrected and flattened spectra (Figure~\ref{fig:spectra}). The number of cut data points and window length were visually found to result in the cleanest fully reduced spectra. Barycentric-corrected wavelengths, errors, and order numbers were taken from the OPERA-reduced data \citep{teeple14} and mapped onto our extracted spectra. Spectra of the same orders from different exposures were interpolated onto the same wavelength grid with three times the original wavelength resolution.

\begin{figure*}[p]
\begin{center}
\includegraphics[width=1.0\textwidth,trim={0.25cm 0 0 0}]{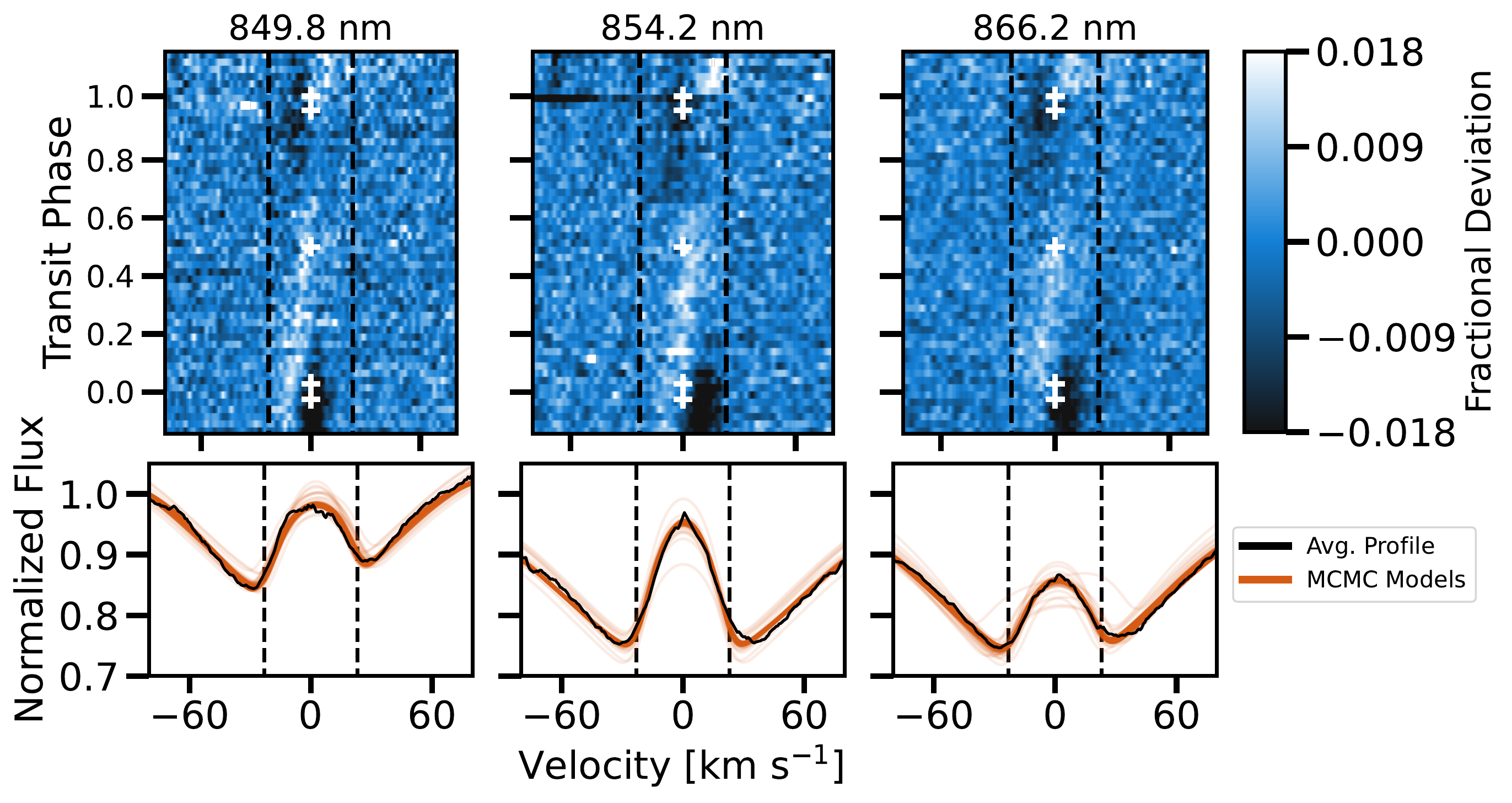}
\caption{Top: Tomographic signal for each line in the \calcium~IRT. Signals are plotted in the rest frame of the star. Excess absorption is shown in white and traces the transit. White pluses represent the four contact points of the transit and transit midpoint. Vertical dashed black lines represent $\pm$ \vsini. Bottom: 200 randomly selected MCMC fits (orange) used to derive the projected obliquity, $\lambda$, of \target, compared to the average line profile (black). Our MCMC best-fit parameters and associated priors are presented in Table~\ref{tab:priors}. \label{fig:ca-waterfalls}}

\vspace{4mm}

\includegraphics[width=1.0\textwidth,trim={0.25cm 0 0 0}]{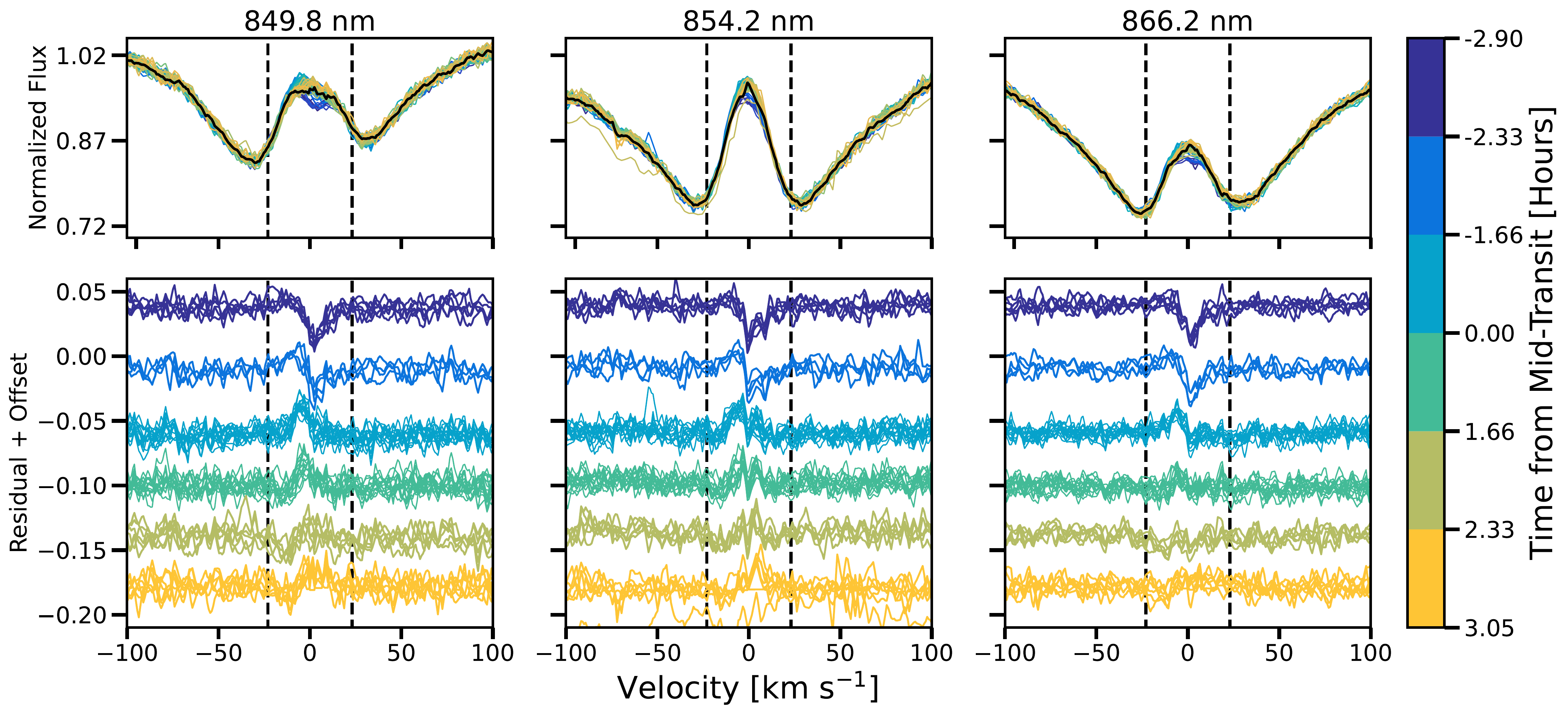}
\caption{Zoom-in of the core and wings of each of the calcium triplet lines. Top: Normalized spectra. The OOT template is plotted as a solid black line. Bottom: The calcium feature with the OOT template subtracted out. There is a deficit in the red side of the core at the beginning of the night (purple), which quickly disappears. An excess in the blue can be seen during the beginning of the transit (dark blue) and slowly disappears through the remainder of the observations (teal to yellow). Lines are binned by time to mid-transit, where the transit duration is 4.66~hours; purple and yellow lines are OOT observations. Vertical dashed black lines represent $\pm$\vsini, centered at the core of each calcium feature. \label{fig:triplet}}
\end{center}
\end{figure*}


\subsubsection{Doppler Tomographic Analysis}\label{subsubsec:tomography}

Doppler tomography relies on a transiting planet creating perturbations in rotationally broadened stellar line profiles that trace the orbit of the planet. Unlike typical radial velocity observations, which rely on tracing changes in the photocenter of lines, the Doppler tomography method allows us to resolve this perturbation in high resolution spectra. Here, we used the Doppler tomographic code \texttt{MISTTBORN}\footnote{\url{https://github.com/captain-exoplanet/misttborn}} \citep{johnson14, johnson17}. In summary, this method compares line profiles between in-transit (IT) and OOT generated models. The initial guess for the line profiles are created by convolving a line list from the Vienna Atomic Line Database \citep{kupka00} with a Gaussian line profile and are rotationally broadened given the \vsini\ of the system. \texttt{MISTTBORN} is flexible, such that after given an initial guess it fits for the line profile and allows it to take on an arbitrary best-fit shape. An average line profile is extracted from each order of each observation and is weighted by the signal-to-noise. The tomographic signal is determined by subtracting the IT from the average line profile across all OOT observations.

\begin{figure}[!ht]
\begin{center}
\includegraphics[width=0.45\textwidth,trim={0.25cm 0 0 0}]{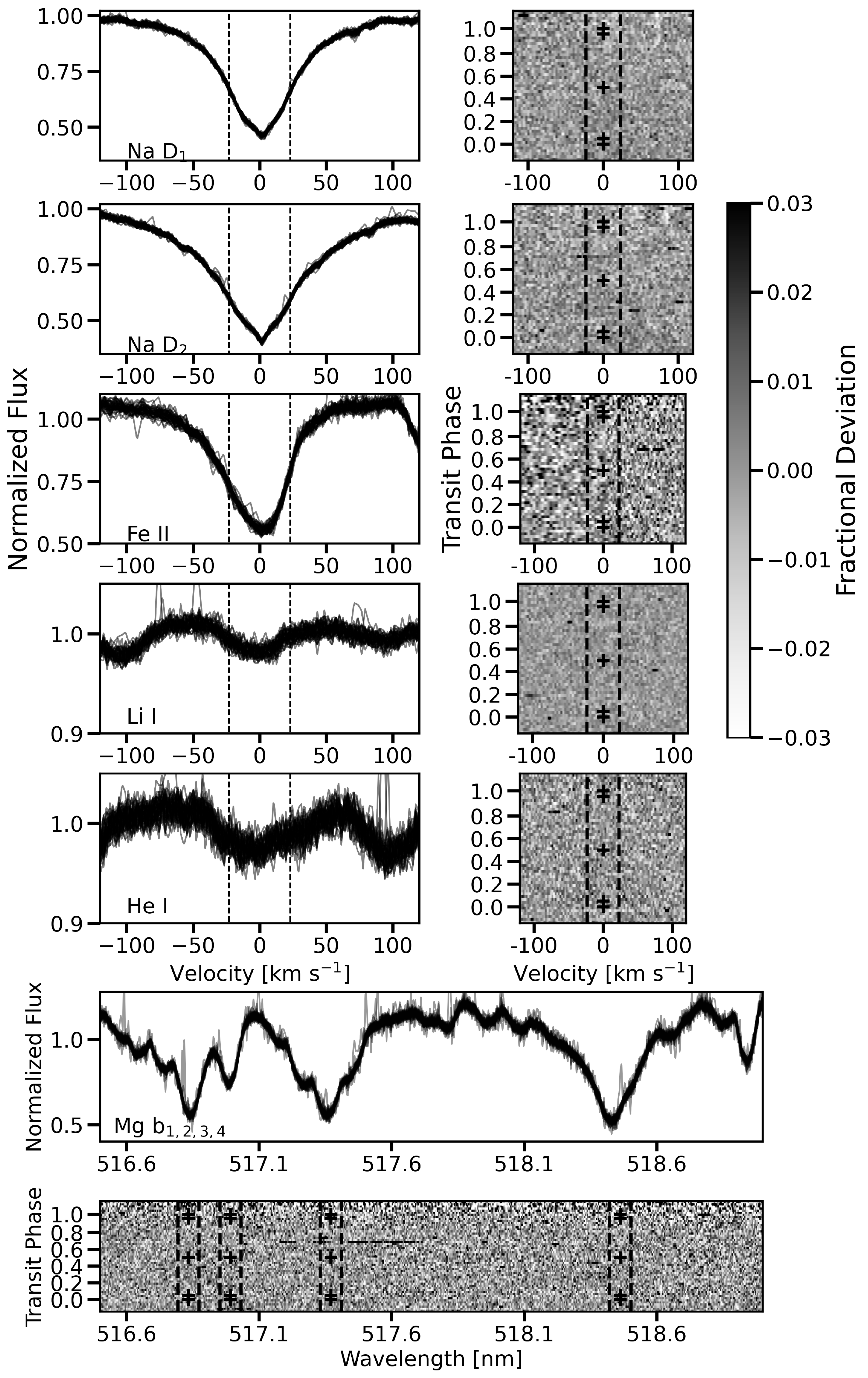}
\caption{Spectral features (line plots) and associated waterfall plots for additional lines where we do not see a Doppler tomographic signal. Each line plot is labeled with the feature and the affiliated waterfall plot is located directly below. We present the Fe\,\doubly\ line at 546.6~nm in the third row, \lithium\ in the fourth row, and \helium\ in the fifth row. The last two rows trace over several nanometers to cover the \magnesium\ b$_{1,2,3,4}$ lines. Lines are colored by time of observation similarly to Figure~\ref{fig:triplet}. Waterfall plots represent deviations from a median OOT template from our observations. White pluses represent the four contact points of the transit and transit midpoint. Vertical dashed black lines represent $\pm$ \vsini. \label{fig:others}}
\end{center}
\end{figure}

In addition to following the methods presented in \cite{johnson14}, we focused on detailed analysis of the Fraunhofer lines \citep{fraunhofer1817}, which trace photospheric activity, and tried to extract the tomographic signal from just these lines. Each feature was compared to a median template created from the OOT observations. Of all the lines, \halpha\ and the \calcium~IRT at 849.8, 854.2, and 866.2~nm were the only lines that showed variation throughout the night. All other lines appeared stable (see examples in Figure~\ref{fig:others}).

\subsection{Veloce-Rosso Observations}

As part of our analysis, we place the \halpha\ variability of \vtau\ in context of other young stars of similar activity levels. During 7 nights between UT 2020-11-08 to 2020-11-26, we observed five young M dwarfs (\teff\ = $3323 - 3584$~K)\footnote{\teff\ taken from the \tess\ Input Catalog v8 \citep[TIC;][]{stassun18}.} with the Veloce-Rosso Spectrograph \citep{veloce_rosso} mounted on the Anglo-Australian Telescope. The targets were originally identified as high-probability ($>94$\%) young moving group candidates using \banyan\ \citep{gagne18}, which uses \Gaia\ kinematics to assign membership, and were presented in \cite{feinstein20}. Three of the targets (TIC 333680372, 246897668, and 178947176) are high-probability kinematic members of the AB Doradus Moving Group ($t_{age} \approx 150$~Myr) and the other two (TICs 146522418 and 427346731) are high-probability kinematic members of the $\beta$~Pictoris Moving Group \citep[\textit{t\textsubscript{age}} $\approx25$~Myr;][]{Bell15}. 

\banyan\ probabilities are strictly based on kinematic arguments and do not guarantee these stars are members of assigned known groups. However, the rapid rotation seen in the \tess\ photometry for these stars is consistent with young ages. Spectroscopic youth indicators, such as strong \halpha\ emission are present in all of our observations, providing further evidence these stars are likely young and potentially members of the AB Doradus and $\beta$ Pictoris moving groups.

Targets were observed for a total of 70-120~minutes per night of observation, with exposure times ranging from 300-600~seconds. Data were taken simultaneously with observations of a laser comb, used for precise wavelength calibration \citep{murphy07}. The laser comb was additionally used to trace individual orders for box extraction. The observed targets have no known transiting planets. We reduced all spectra using methods similar to those described in Section~\ref{subsec:reduction}. The wavelengths were mapped using a pixel-to-wavelength reference made available by the Veloce-Rosso team.\footnote{\url{https://newt.phys.unsw.edu.au/~cgt/Veloce/Veloce.html}} 

Targets were observed simultaneously with \tess. Due to the youth of these stars, it was possible some observations were taken during flare events. The \tess\ 2-minute light curves were searched for flares using the convolutional neural networks developed by \cite{feinstein20}. These neural network models assign a probability that a given event is a flare or not. We considered events ``true flares" when they were assigned a probability $\geq 0.9$. Times of flares were cross-matched with our observation; spectra that were taken during identified flare events were removed from this analysis.

\section{Results}\label{sec:results}

Because we found our spectra to show little variation over the night, we could not use all spectral features to extract a Rossiter-McLaughlin or Doppler tomographic signal. Of the Fraunhofer lines checked, including the \magnesium\ triplet at 517~nm, and Na D\textsubscript{1} and D\textsubscript{2} (Figure~\ref{fig:others}), the tomographic signal was only seen in the \calcium~IRT at 849.8, 854.2, and 866.2~nm. 

\subsection{\calcium\ Infrared Triplet}

\subsubsection{Doppler Tomography \mcmccloud}

The Doppler tomographic analysis of the \calcium~IRT is shown in Figure~\ref{fig:ca-waterfalls}. The fractional deviation compares the difference between the \calcium~IRT lines of each observation with a median profile from all OOT data. Bright areas represent regions of excess absorption, here due to the planet crossing the surface of the star, tracing from the bottom left to top right and approximately from the beginning to end of the transit. The transit signal is seen independently in all three lines. We fit for the tomographic signal in the core of the \calcium~IRT lines using \texttt{MISTTBORN} \citep{johnson14}. The line profile model in \texttt{MISTTBORN} accounts for the distortion of the line shape during the planet's transit and is a function of the parameters presented in Table~\ref{tab:priors}. We do not account for differential rotation or macroturbulence.

\texttt{MISTTBORN} is typically used to model individual lines and not spectral features with cores. Here, we combined the line core with an upside-down Gaussian to model the regions around the core, presented in Figure~\ref{fig:ca-waterfalls}. For \calcium\ at 854.2~nm, we set the central velocity of the Gaussian equal to the fitted line center; the other two Gaussian line centers were allowed to vary due to noticeable asymmetry in the lines. We used \texttt{emcee} \citep{Goodman10, Foreman-Mackey13} to fit each line. We ran our MCMC fit with 350 walkers and 750 steps. After visual inspection, we removed 150 burn-in steps. We verify our chains have converged via visual inspection and following the method of \cite{Geweke92}. The parameters, fit results, and priors are given in Table~\ref{tab:priors}.The median \calcium~IRT and 200 randomly selected MCMC examples are shown in the bottom row of Figure~\ref{fig:triplet}. 

From this, we measure a projected obliquity of $\lambda = $~\obliquity. We find the Doppler tomographic signal is driven by changes in the core of the \calcium~IRT, formed in the stellar chromosphere (Figure~\ref{fig:triplet}). The individual lines (top row) and residuals with a median OOT template (bottom row) are colored by transit phase of \target. The changes in the lines are visible by eye, and are more pronounced in the residuals. There is very little difference between each observation in the wings of the lines (regions outside of $\pm$~\vsini).

\begin{deluxetable*}{l r r r r r}[!ht]
\tabletypesize{\footnotesize}
\tablecaption{Doppler Tomographic Modeling Results for the \calcium~IRT \label{tab:priors}}
\tablehead{\colhead{Planet Parameters} & \colhead{849.8~nm} & \colhead{854.2~nm} & \colhead{866.2~nm} & \colhead{Combined} & \colhead{Prior}\\
\hline \
$\lambda$ [$^\circ$]      & $13.6_{-8.8}^{+16.9}$ & $-0.5_{-22.7}^{+9.3}$ & $4.2_{-13.8}^{+5.8}$ & $4.9_{-15.1}^{+15.0}$ & $\mathcal{U}$[-180, 180]\\
T$_0$ [MJD]               & $58846.09709_{-0.00019}^{+0.00008}$ & $58846.09720_{-0.00018}^{+0.00018}$ & $58846.09716_{-0.00015}^{+0.00017}$ & $58846.09715_{-0.00019}^{+0.00016}$ &$\mathcal{G}$(58846.097156, $10^{-5}$)\\
Period [Days]             & $8.24964_{-0.00012}^{+0.00019}$ & $8.24962_{-0.00011}^{+0.00021}$ & $8.24953_{-0.00020}^{+0.00013}$ & $8.24959_{-0.00018}^{+0.00022}$ &$\mathcal{G}$(8.24958, $10^{-5}$)\\
R$_p$/R$_\star$           &  $0.03923_{-0.00013}^{+0.00034}$ & $0.03926_{-0.00012}^{+0.00063}$ & $0.03869_{-0.00088}^{+0.00052}$ & $0.03921_{-0.00062}^{+0.00033}$ &$\mathcal{G}$(0.039208, $10^{-3}$)\\
\vsini\ [km s$^{-1}$]     & $23.59_{-0.73}^{+0.97}$ & $23.92_{-1.25}^{+0.61}$ & $23.45_{-1.71}^{+0.99}$ & $23.65_{-1.17}^{+0.87}$ &$\mathcal{G}$(23.0, 1.0)\\
a/R$_{\star}$             & $13.188_{-0.136}^{+0.15}$ & $13.187_{-0.005}^{+0.003}$ & $13.218_{-0.088}^{+0.185}$ & $13.189_{-0.047}^{+0.133}$ &$\mathcal{G}$(13.19, 0.01)\\
impact parameter, $b$     & $0.20_{-0.12}^{+0.11}$ & $0.21_{-0.09}^{+0.12}$ & $0.22_{-0.09}^{+0.10}$ & $0.21_{-0.10}^{+0.11}$ &$\mathcal{G}$(0.20, 0.05)\\
eccentricity, $e$         & $0.11_{-0.01}^{+0.03}$ & $0.09_{-0.05}^{+0.02}$ & $0.09_{-0.04}^{+0.03}$ & $0.10_{-0.04}^{+0.03}$ &$\mathcal{U}$[0, 0.5]\\
periastron, $\omega$ [$^\circ$]                 &  $90.18_{-65.7}^{+59.52}$ & $87.27_{-63.2}^{+59.61}$ & $86.4_{-56.23}^{+51.07}$ & $87.84_{-61.78}^{+57.05}$ &$\mathcal{U}$(-180, 180)\\
limb darkening, $u_1$     & $0.41_{-0.10}^{+0.21}$ & $0.86_{-0.49}^{+0.13}$ & $0.63_{-0.26}^{+0.25}$ & $0.55_{-0.19}^{+0.38}$ &$\mathcal{U}$[0, 1]\\
limb darkening, $u_2$     & $0.21_{-0.14}^{+0.12}$ & $0.30_{-0.15}^{+0.21}$ & $0.35_{-0.09}^{+0.20}$ & $0.30_{-0.17}^{+0.19}$  &$\mathcal{U}$[0, 1]\\
line center [km s$^{-1}$] & $-1.31_{-0.68}^{+0.02}$ & $-7.29_{-0.02}^{+0.19}$ & $-14.72_{-0.11}^{+0.67}$ & --- &$\mathcal{U}$[-20, 20]\\
\hline\
Gaussian Parameters\\
\hline\
mean, $\mu$, [km s$^{-1}$] & $4.99_{-0.26}^{+0.19}$ & --- & $-12.57_{-0.99}^{+1.55}$ & --- &$\mathcal{U}$[-20, 20]\\
width, $\sigma$     & $39.36_{-0.72}^{+3.55}$ & $57.66_{-2.37}^{+1.43}$ & $53.30_{-1.75}^{+1.74}$ & --- &$\mathcal{U}$[35, 60]\\
scaling factor      & $23.56_{-0.38}^{+1.62}$ & $41.53_{-2.71}^{+1.92}$ & $36.99_{-1.68}^{+2.28}$ & --- &$\mathcal{U}$[10, 50]\\
core scaling factor & $0.176_{-0.012}^{+0.005}$ & $0.24_{-0.003}^{+0.003}$ & $0.145_{-0.004}^{+0.003}$ & --- &$\mathcal{U}$(0, 1]\\
model y-offset      & $1.04_{-0.0}^{+0.01}$ & $1.00_{-0.01}^{+0.01}$ & $0.99_{-0.01}^{+0.01}$ & --- &$\mathcal{U}$[0.5, 1.5]
}
\startdata
\enddata
\tablecomments{\calcium\ at 852.4~nm was symmetric about the core so the mean of the underlying Gaussian, $\mu$, was not fit separately from the line center.}
\end{deluxetable*}

\begin{figure}[!ht]
\begin{center}
\includegraphics[width=0.4\textwidth,trim={0.25cm 0 0 0}]{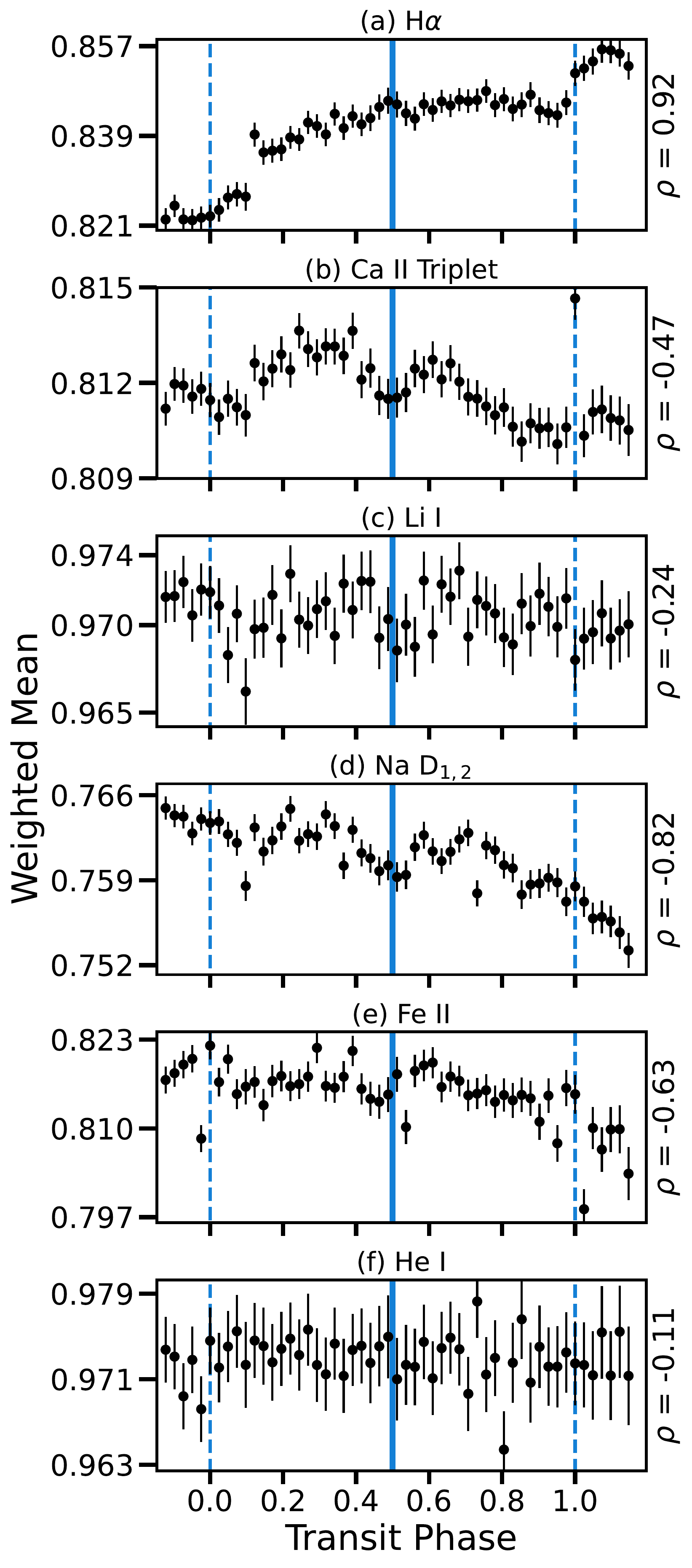}
\caption{Weighted mean ``light curves", $\overline{x}$, and affiliated errors, $\sigma_{\overline{x}}$, of the (a) \halpha, (b) \calcium~IRT, (c) \lithium\ at 670.7~nm, (d) the \sodium\ doublet at 589~nm, (e) Fe~\doubly at 546.6~nm, and (f) \helium\ at 587.8~nm. The Spearman's correlation value, $\rho$, for each feature is presented on the right hand side. Note the scale for each subplot was chosen to optimize the entire region and error bars are comparable. The solid line marks $t_{mid}$ and the dashed lines represent $t_1$ and $t_4$. An increase in weighted mean corresponds to excess absorption in the spectral feature. There is a visible increase in \calcium\ with transit ingress and egress, while no such trend is seen in the \lithium. This provides additional confidence the deviations in \calcium\ is planetary in nature. The spectra for additional lines are presented in Figure~\ref{fig:others}. \label{fig:weighted-means}}
\end{center}
\end{figure}

\begin{figure*}[!ht]
\begin{center}
\includegraphics[width=1.0\textwidth,trim={0.25cm 0 0 0}]{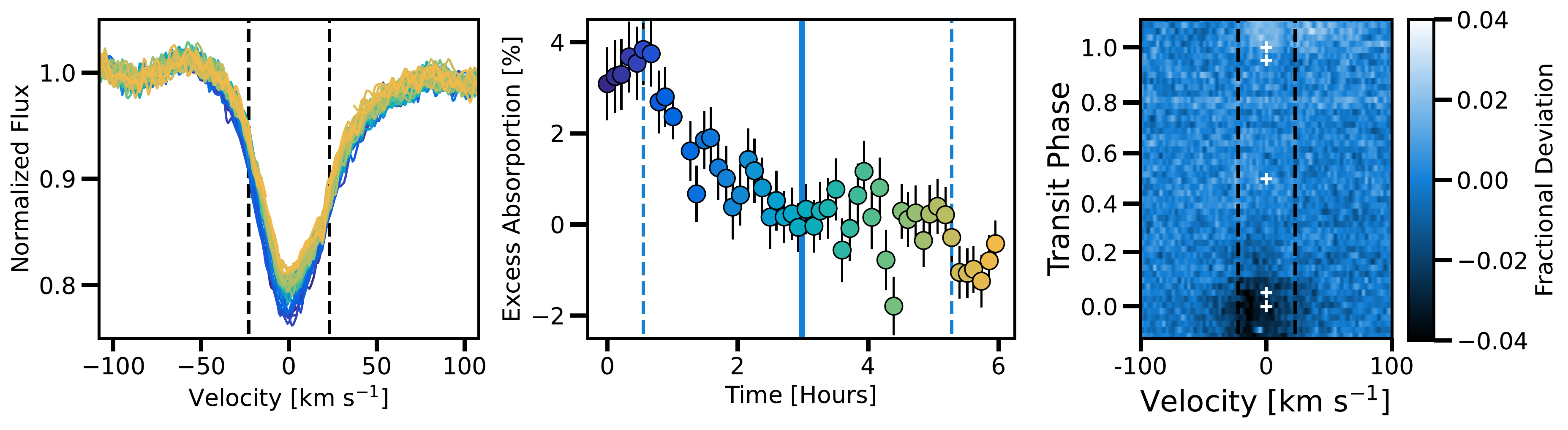}
\caption{Demonstration of variation in \halpha\ observed in our observations. Left: Normalized spectra colored by observation time. Purple and yellow represent the beginning and end of the night, respectively. The smooth variation in depth of the \halpha\ feature is clear. Middle: Measurement of excess absorption seen in \halpha\ as a function of time. The color of points correspond to color of the spectra in the left plot. The solid line corresponds to $t_{mid}$. The dashed lines correspond to transit contact points $t_1$ and $t_4$. Right: Tomography of \halpha\ colored by fractional deviation from the OOT template. White pluses represent the four contact points of the transit and transit midpoint. Vertical dashed black lines in the left and right panels represent $\pm$ \vsini. \label{fig:halpha}}
\end{center}
\end{figure*}

\subsubsection{Weighted Mean ``Light Curves" \ewcloud}

Comparing the weighted contributions of spectral features of IT to OOT observations has been used to infer the presence of planetary atmospheres, in particular the \sodium\,D lines \citep{charbonneau02, Wyttenbach15}. For this, we follow a similar approach to \citep{barnes16}. Per each spectral feature, we calculate:

\begin{equation}
    x_{i}(F_j, \sigma_j^2) = \sum_{j=0}^m (F_{j} / \sigma_{j}^2) \,\,\, /\,\,\, \sum_{j=0}^m (1 / \sigma_{j}^2) 
\end{equation}

where $x_i$ refers to a specific spectral feature, $F_j$ is the flux array and $\sigma_j^2$ is the variance array for $j$ observations. The total weighted mean ``light curve" across multiple spectral features is then given by

\begin{equation}
    \overline{x}(F_j, \sigma_j^2) = \sum_{i=0}^n x_{i}
\end{equation}

where $n$ is the total number of combined features (e.g. for the \calcium~IRT, $n=3$). In a similar manner, such that the associated standard deviation for a given feature is:

\begin{equation}
    \sigma_i (\sigma_j^2) = \sum_{j=0}^m \frac{1}{\sigma_j^2}
\end{equation}

where $\sigma_i$ refers to errors for a specific spectral line and $\sigma_j^2$ is the variance array for $j$ observations. The total weighted standard deviation across $i$ spectral features is calculated as:

\begin{equation}
    \sigma_{\overline{x}} (\sigma_j^2) =  \left( \sum_{i=0}^n \sigma_i \right)^{-1/2}
\end{equation}

We compare a weighted mean of the the \calcium~IRT lines to various other lines and activity indicators, such as \lithium\ at 670.7~nm, the \sodium\ doublet at 589~nm, and the \helium\ line at 587.8~nm; the results are shown in Figure~\ref{fig:weighted-means}. An increase in weighted mean can be interpreted as excess absorption in the spectral feature.

\lithium\ has been seen to remain relatively stable in the presence of high levels of activity and photometric spot variations \citep{pallavicini92, pallavicini93, soderblom93}, with a potential abundance increase in the presence of the more-spotted hemisphere \citep{flores-soriano15}. We measured the \lithium\ equivalent width, and found the average to be 140~m\AA, which is consistent with other K0-K1.5 stars of this age \citep{wichmann00}. As a comparison, we use \lithium\ as a control to what is intrinsically occurring in the stellar chromosphere. We test for correlations between transit phase and the light curves for \lithium\ and \calcium~IRT in Figure~\ref{fig:weighted-means} by calculating the Spearman's rank correlation coefficient \citep{spearman07}. This test assess how well two data sets can be described using a monotonic function. A correlation coefficient of $\pm 1$ implies an exact monotonic relationship, while 0 implies no correlation. We find correlation values of -0.24 and -0.65 for \lithium\ and \calcium~IRT respectively, implying \calcium~IRT is slightly more correlated with transit phase than \lithium. \lithium\ could be enhanced from flare events; however, we see no evidence of such event in other spectroscopic features affected by flares \citep[e.g. variations in the blue wing of \halpha;][]{maehara21}. 

\subsubsection{Correlations Between Spectral Features}

We evaluate other known activity indicators, such as \sodium\ doublet, Fe~\doubly, and \helium\ (subplots d - f in Figure~\ref{fig:weighted-means}). We calculated the Spearman's correlation value for each to be -0.80, -0.63, and -0.11, respectively. High Spearman's correlation values could be due to long term stellar activity trends seen in the data. This could be particularly true for the \sodium\ doublet. Although trends in the weighted mean is similar to \calcium~IRT, there is no evidence of a Doppler tomographic signal in the lines (Figure~\ref{fig:others}). For completeness, we also evaluate the Spearman's correlation value for \halpha\ (subplot a in Figure~\ref{fig:weighted-means}) and find it to be 0.93, indicating a strong correlation with transit phase. We further inspect the \halpha\ signal below.

\begin{figure}[!ht]
\begin{center}
\includegraphics[width=0.44\textwidth,trim={0.25cm 0 0 0}]{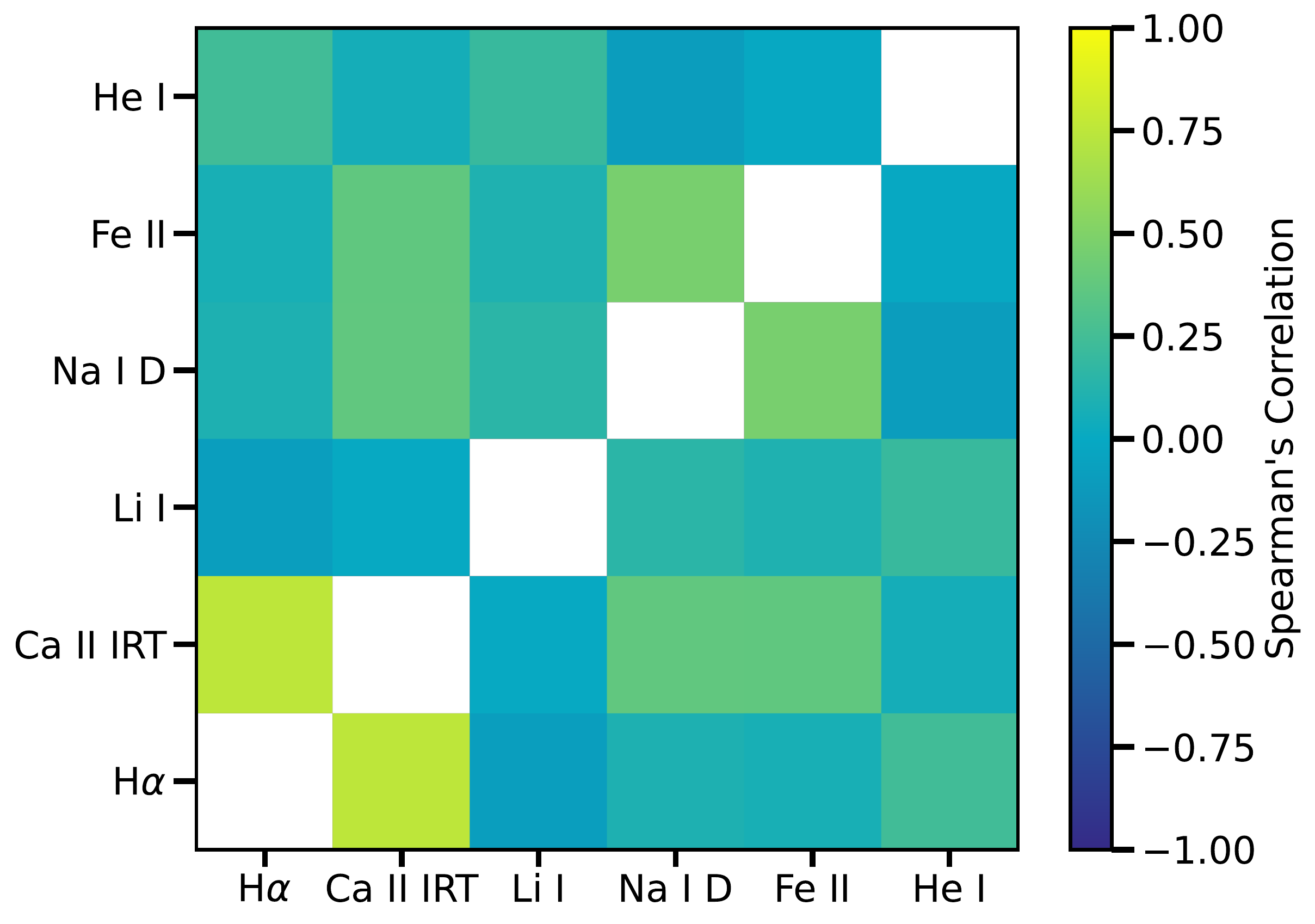}
\caption{The Spearman's correlation value for spectral features presented in Figure~\ref{fig:weighted-means}. The average Spearman's value across all spectral features is 0.1, indicating that there is no correlation.The strongest correlation exists between \halpha\ and the \calcium~IRT.  \label{fig:spearmans}}
\end{center}
\end{figure}

We additionally evaluate the Spearman's correlation value between each of the spectral features in Figure~\ref{fig:weighted-means}. In order to look for correlations in the features, rather than long term trends, we fit a first degree polynomial to each weighted mean light curve. From the trend-removed light curve, we found the correlation between each line. The correlation values are shown in Figure~\ref{fig:spearmans}. The median correlation value between all the lines is 0.1, which indicates there is very little to no correlation. The lines that are most strongly correlated are \halpha\ and the \calcium~IRT. Once the long term trend is removed from the \halpha, there is a clear upwards ``bump", similar to what is seen in Figure~\ref{fig:halpha}. Due to the strong correlation between that ``bump" and the increase that is seen in the \calcium~IRT, which is planetary in nature, this could indicate that the trend in \halpha\ is also originating from \target.

\subsection{\halpha\ Variations}

The \halpha\ variations are shown in Figure~\ref{fig:halpha}. The leftmost panel is the \halpha\ line and surrounding continuum used in our analysis. By eye, there is a clear trend in decreasing absorption throughout the night. To quantify this trend, we calculate the equivalent absorption (EA) of each line, defined as:

\begin{equation}\label{eq:ew}
    \textrm{EA} = - \sum_{i=0}^n \frac{F_i}{F_{out}} - 1
\end{equation}

where $F_i$ is the line flux for one observation and $F_{out}$ is the median template of the OOT observations. The equivalent absorption for \halpha\ is shown in the middle panel of Figure~\ref{fig:halpha}. There is a trend of decreasing excess absorption throughout the night, with potentially sharp features after \target\ is fully visible across the stellar surface. This could be the presence of an extended hydrogen atmosphere. However, given the long term trend and minimal OOT observations, it is challenging to disentangle this signature from general stellar activity.

The Doppler tomographic analysis is shown in the rightmost panel of Figure~\ref{fig:halpha}. Dark and light regions at the beginning and end of the night are the result of over-subtraction from the template, which is constructed from these specific observations. There is no clear tomographic signal seen between $\pm$~\vsini, as was seen in Figure~\ref{fig:ca-waterfalls}. This is likely due to stellar activity dominating the long-term trend in \halpha\ seen over the night. The excess absorption seen in the middle panel is therefore most likely stellar activity.

\subsubsection{Comparison to Veloce-Rosso Targets}

\begin{figure}[!ht]
\begin{center}
\includegraphics[width=0.46\textwidth,trim={0.25cm 0 0 0}]{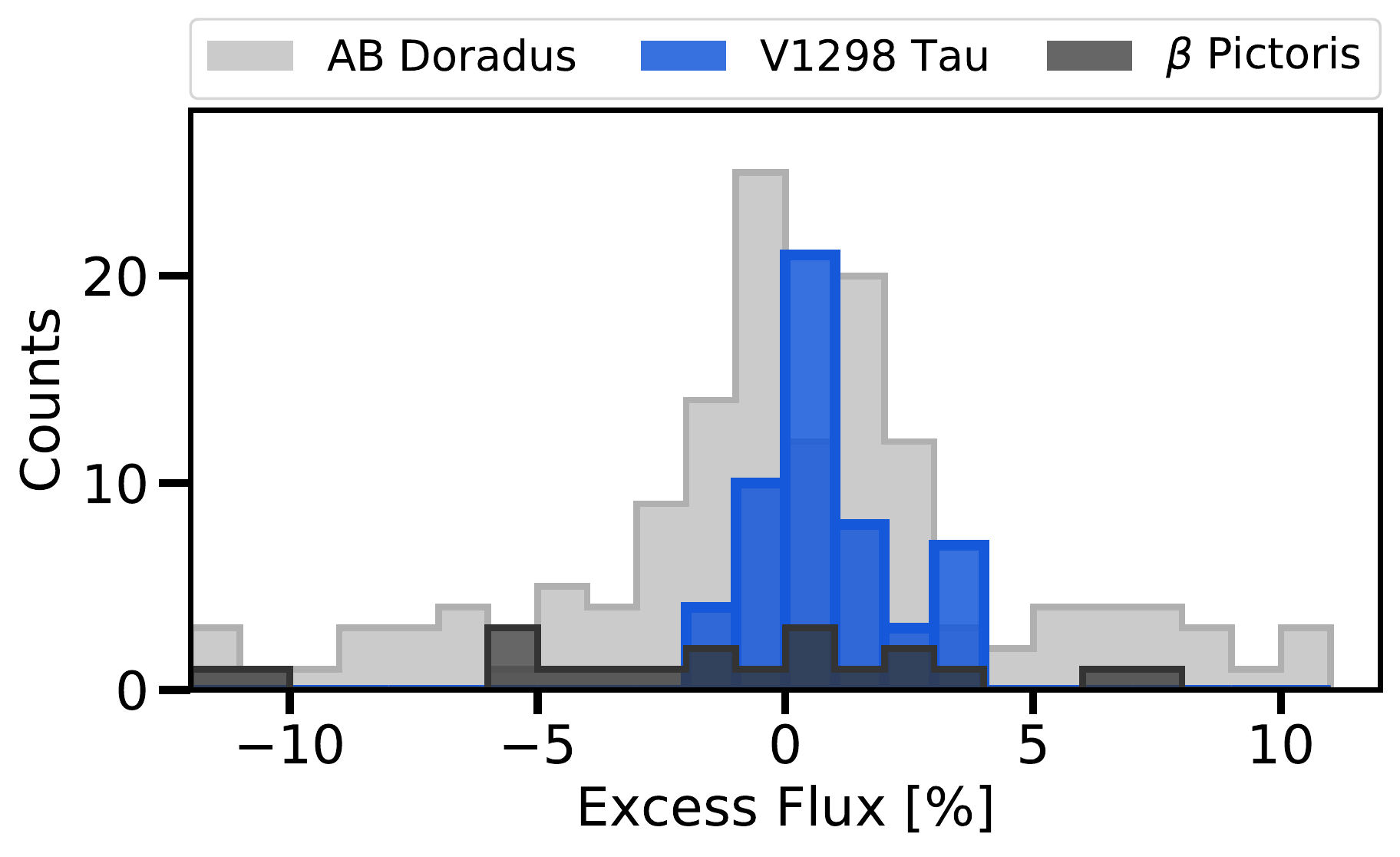}
\caption{Comparison of \halpha\ variations observed in \vtau\ (purple) with five young M stars observed with Veloce-Rosso. The levels of \halpha\ emission in these young stars varies on the order of a few percent over a few hours. The \halpha\ for \vtau\ is correlated with time, hinting this may be the signature of an escaping hydrogen atmosphere. However, observations of other young active stars show similar trends over a single night. The \halpha\ excess variations of \vtau\ trace the distribution of AB Doradus members. This is likely because the relative activity of a 23~Myr K star is comparable to a 150~Myr M star. The light gray histogram are members of AB Doradus moving group ($t_{age}=150$~Myr); dark the gray histogram dark gray points are members of the $\beta$~Pictoris moving group ($t_{age} = 25$~Myr). Veloce-Rosso observations during flare events were removed. Each bin represents 1\% variation in excess flux. \label{fig:veloce}}
\end{center}
\end{figure}

The EA of \halpha\ was additionally calculated for the young stars in our Veloce-Rosso sample using Equation~\ref{eq:ew}.  We note that the activity of M and K stars are comparable at ages $\leq 1$~Gyr \citep{schneider18, richey-yowell19}. Thus, this comparison will help provide a stellar context for the \halpha\ variations. The results are shown in Figure~\ref{fig:veloce}. We note that the \halpha\ for targets observed with Veloce-Rosso is in emission, while \vtau\ clearly shows \halpha\ in absorption without a significant core component.

A median template of \halpha\ was created per star per night of observation. The excess flux variation is calculated by comparing each exposure to the median template. This is similar to the way excess \halpha\ flux for \vtau\ was calculated. There is clear variation for all young stars, with excess variation extending out to $\pm 10$\%. The predominant amount of excess flux varies between $\pm 3$\%, which follows the trend seen for \vtau. We additionally found stars in our sample exhibit smooth increasing or decreasing \halpha\ variation over a single night, in a similar way to \vtau\ (Figure~\ref{fig:halpha}, middle panel). This demonstrates the behavior of \halpha\ in young stars can change dramatically over a night. 

The average strength \halpha\ emission for young stars is comparable to what is seen in our Veloce-Rosso sample. AB Doradus and other confirmed young moving group members have \halpha\ EWs ranging from $-10$ to $0$, where negative values indicate emission \citep{riedel17}. Confirmed $\beta$~Pictoris moving group members were seen to have \halpha\ EWs ranging from $\sim -15$ to $0$ for M0-M9 stars \citep{Shkolnik17}. \cite{schneider19} measured \halpha\ EWs in 336 young low-mass stars (K5-M9; $t_{\textrm{age}} \sim 22 - 200$~Myr) with values ranging from $\sim -20$ to $2$, where earlier spectral types tend towards positive EWs. The values for our \halpha\ EA measurements fall well within the ranges for other young stars with ages between $t_{\textrm{age}} \sim 23 - 150$~Myr. In the context of other young stars, the variations seen in \vtau\ can be solely attributed to youth and general stellar activity.

\section{Discussion}\label{sec:discussion}

We speculate on several different explanations for the observed \halpha\ variations, such as limb darkening, the presence of star spots, and the presence of an extended \halpha\ atmosphere from \target\ or \vtau\,d. 

\subsection{Center-to-Limb Variations \clvcloud}\label{subsec:clv}

The underlying wavelength dependence of limb darkening can create signals masquerading as those expected from planet atmospheres \citep{Czesla15}. We model center-to-limb variations (CLVs) to address this possibility. A synthetic spectrum was obtained from the \cite{castelli94} grid of stellar atmospheric models generated with \texttt{ATLAS9} \citep{kurucz93}. Based on stellar parameters presented in \cite{david19_v1298b}, we chose a model with T$_{\textrm{eff}} = 5000$~K and \logg\,=\,4.00 with solar metallicity. Spectra with different limb darkening values were generated using \texttt{spectrum} \citep{gray94} with a wavelength step size of $\Delta \lambda = 0.002$\angstrom. The spectra were additionally rotationally broadened  (\vsini\ $\sim 23$~km s$^{-1}$) following the formalism of \cite{gray08}. A total of 50 synthetic spectra were generated from $u = 0 - 1$, where $u = 0$ is the edge of the stellar disk and $u = 1$ is the center. We sampled limb angles at $\Delta u = 0.02$; $u = 0.0$ was replaced with $u = 0.001$ \citep{Czesla15}.

\begin{figure}[!ht]
\begin{center}
\includegraphics[width=0.47\textwidth,trim={0.25cm 0 0 0}]{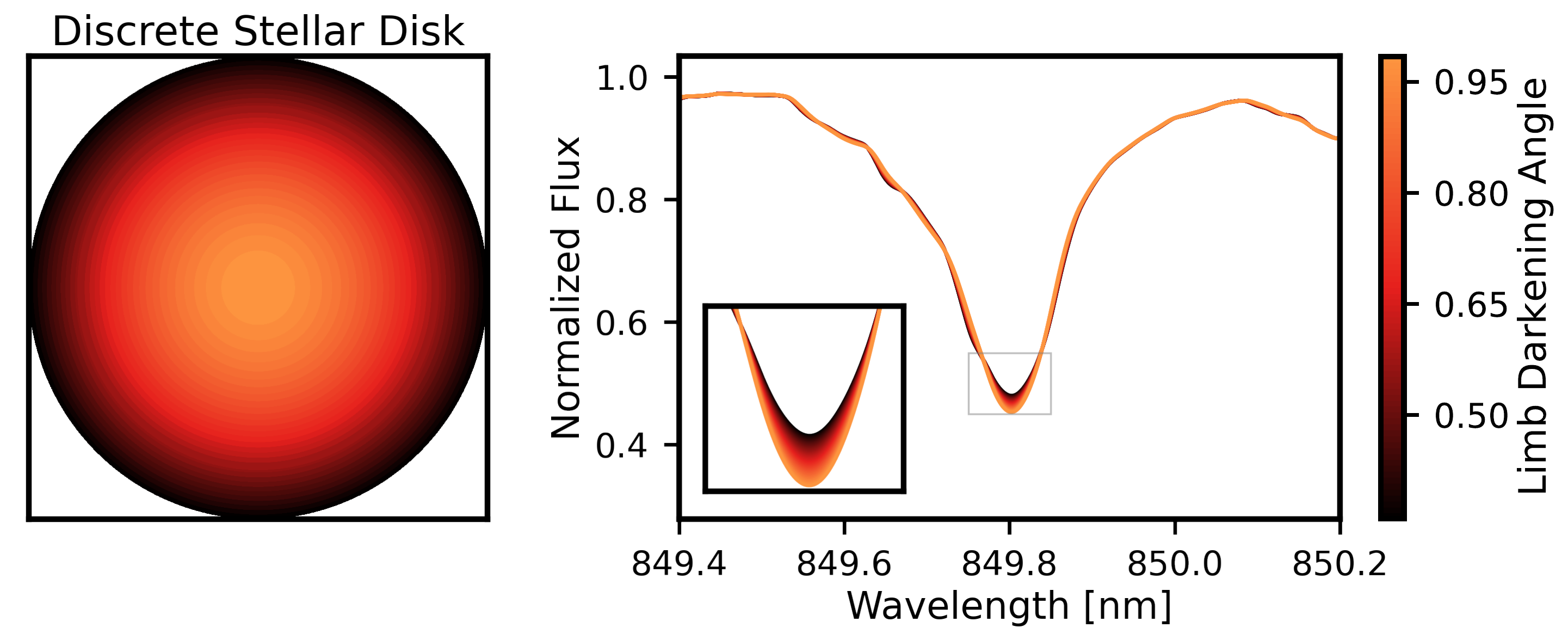}
\caption{Modeled stellar disk and accompanying spectra for \vtau\ at different limb darkening angles. The original spectrum was taken from the grid of stellar atmospheric models from \cite{KuruczModels}. Spectra were then generated at different limb darkening angles using \texttt{spectrum}. We accounted for rotational broadening due to high \vsini of \vtau. The averaged affect of limb darkening over the entire stellar surface is negligible with respect to the \halpha\ variation seen during our observations. \label{fig:clv}}
\end{center}
\end{figure}

We created a discretized stellar surface with $R_{pixels} = 20001$, filled with concentric circles for each value of $u$ as prescribed by \cite{vogt87}. An example of the modeled stellar disk and accompanying spectral feature for one of the \calcium~IRT lines at 849.8~nm is shown in Figure~\ref{fig:clv}. Variations in line depth as a function of $u$ are highlighted in the inset panel. We model the transit of \target\ with a radius of 19 pixels, maintaining an $R_p/R_\star = 0.03$ and at an impact parameter of $b = 0.2$ \citep{david19_v1298all}. The spectrum at each step is the summation of all unocculted elements. 

We find that CLVs for this system induce changes in the spectral features on the order of $10^{-5}$. These effects are negligible compared to the observed 8~ppt signal (Figure~\ref{fig:weighted-means}). The CLV testing described above does not account for non-local thermodynamic equilibrium effects. Although it would be more rigorous to include such calculations, the magnitude of the variations is negligible compared to the contribution of ARs \citep{cauley18}. Therefore, the behavior seen here is not attributable to CLV. 

Next, we account for potential limb-brightening in the core of the Ca II IRT lines, following the methods presented in \cite{Czesla15}. First, we create a difference light curve ($DC$) with a modified version of Equation 9 to account for three, instead of two, lines:

\begin{equation}
    DC_\textrm{core}(t_i) = \frac{1}{3} \sum_{j=0}^3 LC_j(t_i) - \frac{1}{3} \sum_{j=0}^3 LC_{C_j}(t_i)
\end{equation}

where $t_i$ is an observation, $LC_j$ is the mean weighted light curve of each of the three individual \calcium~IRT line, $j$, and $LC_{C_j}$ is the mean weighted light curve for the continuum around each \calcium~IRT line. Due to there being many lines near the \calcium~IRT, we select the continuum region as 0.5~nm on either side of the lines. We find the effect of limb brightening to be on a similar scale to that in \cite{Czesla15}, but does not exhibit the same trend. While limb brightening could be responsible for this signal, it is near-impossible to distinguish between variability from the planet and CLV-induced changes, and therefore we rule out limb-brightening as the source of this variability.


\subsection{Stellar Nature of the Variability \spotcloud}\label{subsec:spots}

Due to \halpha\ being a tracer of stellar activity, we must explore the possibility that the observed excess absorption variations are strictly stellar in nature. The smooth variation over our observations was seen in similar observations for targets in our Veloce-Rosso sample. Additionally, young stars are believed to have significant photospheric inhomogeneities, seen in the original light curve of \vtau. Similar night-long trends in activity have been seen for other young planet hosts \citep{montet20} and the trend in the excess \halpha\ absorption during our observations can be explained by stellar activity alone.

\subsubsection{Spot \& Faculae Modeling}

We use \starry\ \citep{Luger19}, an analytic solution to time series of planet transits and stellar surfaces based on applications of spherical harmonics, in combination with \texttt{spectrum} to recreate the \halpha\ absorption with starspots and surrounding facular regions. We follow the methods discussed in Section~\ref{subsec:clv} and add in the presence of two starspots with surrounding facular regions. We take a stellar spectrum model of the same \logg\ and metallicity as \vtau\ at 5500~K and 4500~K for the faculae and spots respectively. Although these \teff\ may not accurately represent all starspot/facular regions, they provide a reasonable estimation for determining changes in line depths/profiles in the presence of such features \citep{catalano2002}. 

\begin{figure}[!ht]
\begin{center}
\includegraphics[width=0.45\textwidth,trim={0.25cm 0 0 0}]{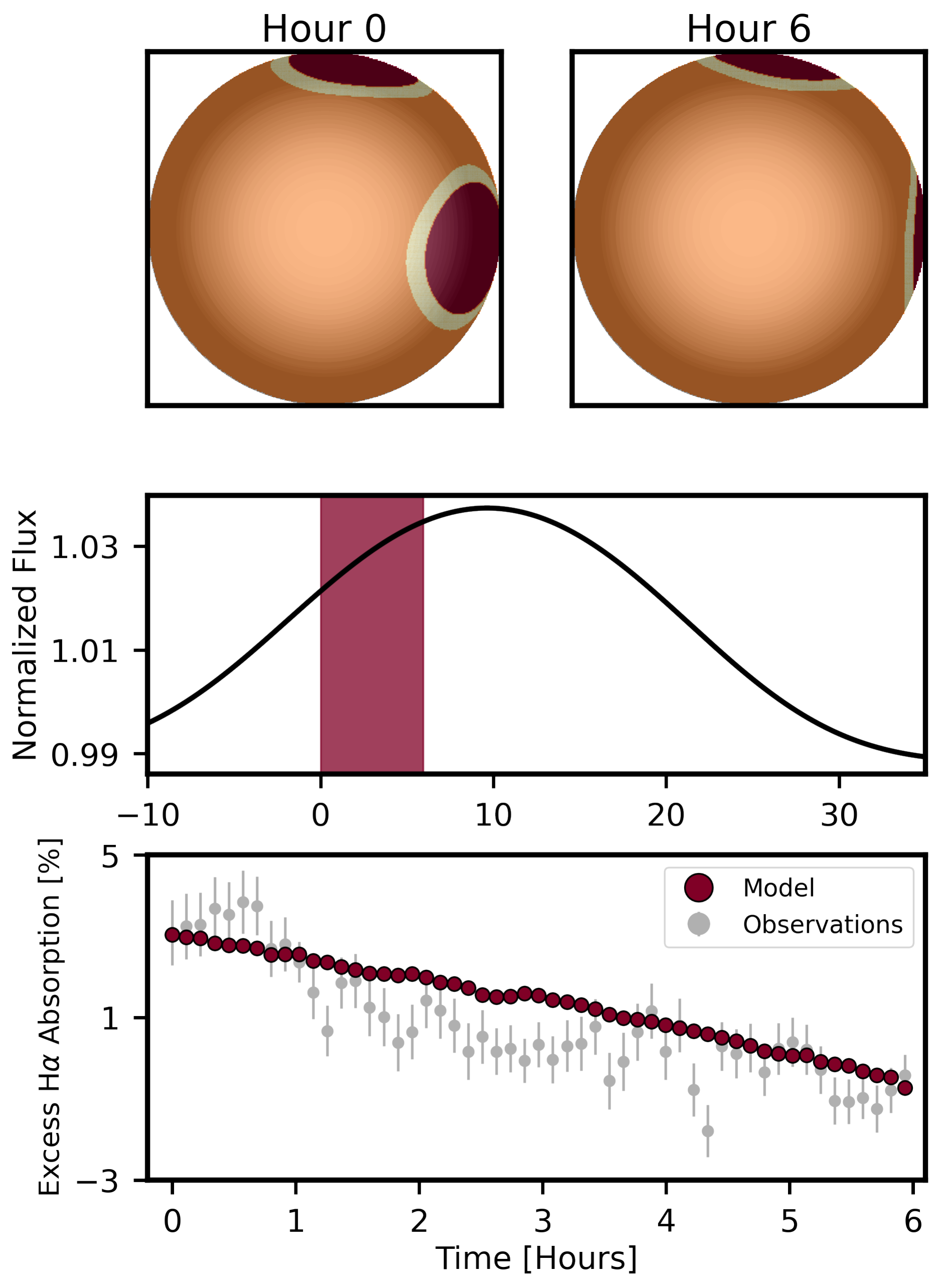}
\caption{Toy spot and facula model created using \starry\ of a potential configuration on \vtau\ that produces similar levels of \halpha\ variation to that seen in our observations. Top row: Surface maps at the beginning and after 6 hours of rotation. Dark red regions represent starpots; bright yellow represent surrounding facular regions; orange is the surface. Maps are limb-darkened. The starspots and faculae comprise of $\sim 20\%$ of the visible surface at the start. Middle row: Normalized light curve of this starspot configuration. Variation amplitudes match those that are present in the original \ktwo\ observations \citep{david19_v1298b}. Shaded red region represents the 6 hour window for the above rotation configurations. Bottom row: Measured \halpha\ excess absorption at each rotation time step. Identifying individual starspots is a degenerate issue; this is one configuration that is able to reproduce a similar trend in \halpha\ excess absorption over the timescale of our observations. \label{fig:starry}}
\end{center}
\end{figure}

We created \starry\ models as a function of rotation that are 2500 $\times$ 2500 pixels in size. Then we discretized the surface to indicate values of either surface (orange), starspot (red), or facula (yellow); examples are shown in the top row of Figure~\ref{fig:starry}. The approximate area of spot to facula is 60\%, similar to that in \cite{cauley17} and the associated modeled photometric variability (Figure~\ref{fig:starry}, middle) corresponds to variability seen in the original \ktwo\ light curve \citep{david19_v1298b}. Forward modeling of \ktwo/\tess\ light curves also suggests the variability is driven more strongly by starspots than faculae \citep{johnson21}. We weighted the spectra by the area of each feature (spot, facula, or surface) and accounted for limb-darkening.

The configuration explored in our toy model is just one of many potential solutions and was not fit to the original \ktwo\ light curve, but rather captures a general scale of spot variability. Therefore, it should not be taken as the ground-truth configuration \citep{luger21}. A six-hour window is highlighted in the photometric variability in red (Figure~\ref{fig:starry}, middle), which corresponds to the \halpha\ excess absorption seen in the bottom row of the same figure. Our observations (gray) show a decrease in excess \halpha\ absorption of 4.29\% over six hours; this model (red) has a similar trend, decreasing by 4.37\% over the simulation.

\subsection{Planetary Nature of the Variability}

We speculate on several different planetary configurations that may be the source of the \halpha\ and \calcium~IRT variations seen over our observations. Each scenario is evaluated separately, although the ground truth could be the result of some combination from each. 

\subsubsection{Extended \halpha\ Atmosphere of \vtau\,c}\label{subsubsec:c_atm}

Given the age of the system and current X-ray irradiation, \target\ is predicted to evolve significantly from $5.59 R_\oplus$ to a final radius of $1-5.55 R_\oplus$ given the different mass and X-ray activity level assumptions \citep{poppenhaeger20}. It would therefore not be surprising to detect a highly extended atmosphere from our target. We note that slight ``bumps" in the measured \halpha\ equivalent width at transit ingress and egress for \target\ (Figure~\ref{fig:halpha}; middle panel) could be attributed to the planet, while the overall slope is dominated by stellar activity. We fit a line between the equivalent widths for ingress and egress to represent the activity longer-term trend and measured a depth in the excess absorption of $1.85\%$. Assuming spherical symmetry, this would correspond to an atmosphere with a thickness of 1.28~R\textsubscript{J}.

\subsubsection{\halpha\ Tail from \vtau\,d}\label{subsubsec:d_atm}

With updated \textit{Spitzer} ephemerides, we also find that the transit of \vtau\,d ended $12 \pm 15$~minutes before our observations \livingstoncitep. \vtau\,d is a $6.41 R_\oplus$ planet on a 12.4 day period and is predicted to evolve to $1.5-6.4 R_\oplus$ in 5~Gyr  \citep{poppenhaeger20}. If an atmospheric tail is present from \vtau\,d, it could be contaminating our observations at the beginning of the night. The first seven observations ($t_{\textrm{MJD}} = 58870.725 - 58870.753$) show a relatively stable \halpha\ excess absorption of 5\% before decreasing. This could very well mean that our initial observations are catching the tail end of an inflated hydrogen atmosphere around planet\,d. 

In this scenario, we can estimate the extent of the hydrogen atmosphere. If we assume the transit of the extended atmosphere continues for the first 2.5 hours, up until the excess \halpha\ absorption temporarily flattens out, we see the signal change $\Delta 5$\%. A 5\% signal would correspond to an atmosphere thickness of 2.36~R\textsubscript{J}. Assuming the atmospheric signal continues for 2.5~hours, this would correspond to a tail length of 5.85~R\textsubscript{J}. The remaining observations would be affected by the transit of \target, with a potential active region crossing resulting in increased noise in our \halpha\ measured EAs between $t_{\textrm{MJD}} = 58870.871 - 58870.908$ (corresponding to 3.504-4.392 hours in the middle panel of Figure~\ref{fig:halpha}).

\subsubsection{\calcium~IRT from \target}\label{subsubsec:spi}

\begin{figure*}[!ht]
\begin{center}
\includegraphics[width=\textwidth,trim={0.25cm 0 0 0}]{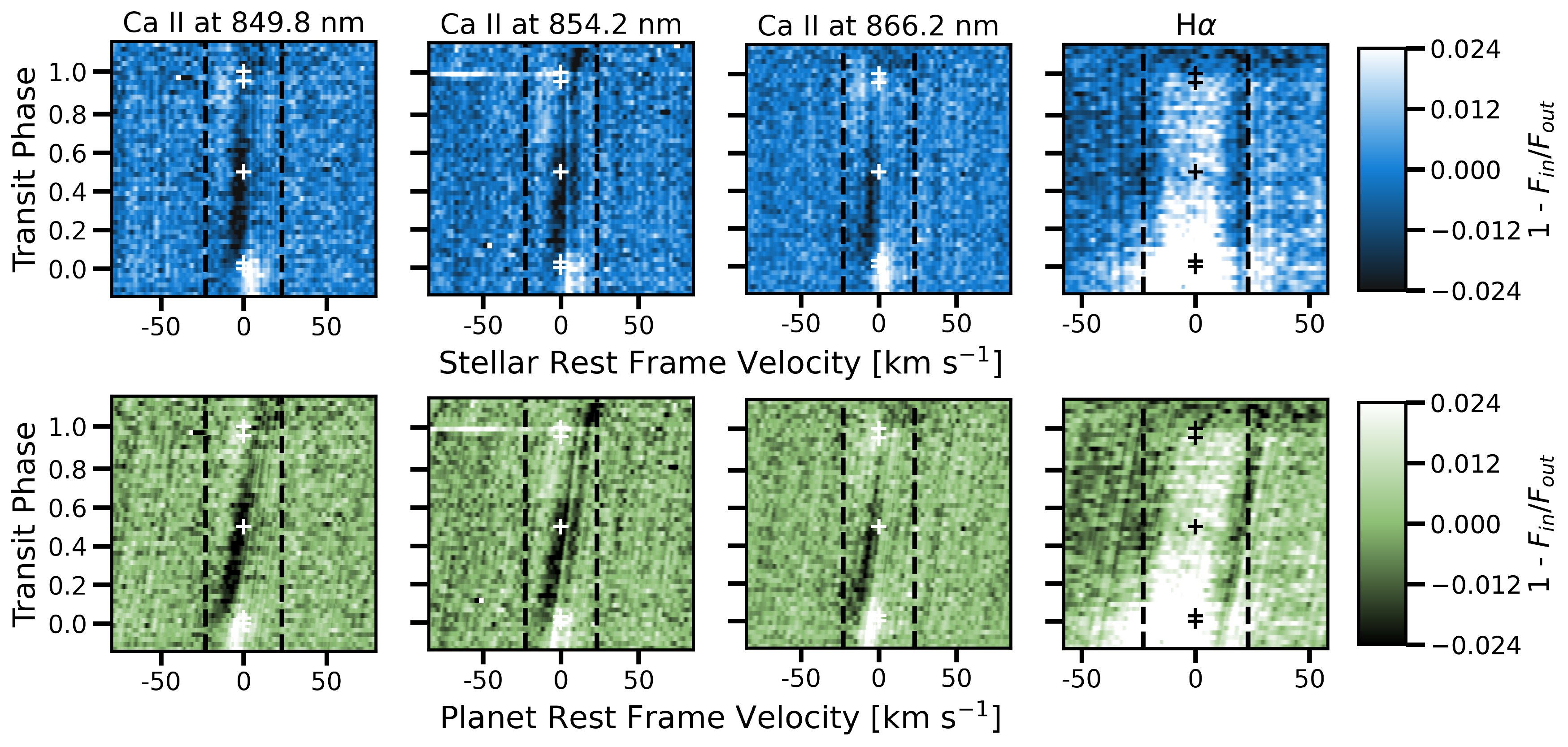}
\caption{Comparison of transmission signal for both \calcium~IRT and \halpha\ in the stellar rest frame (top; blue) and \target's rest frame (bottom; green). Here, white represents absorption and black represents emission, which is not physical in transmission spectra. The excess in \calcium~IRT stretches more so from $\pm$ \vsini\ (vertical dashed black lines) in the planet's rest frame velocity, indicating the signal is stellar in origin. Pluses represent the four contact points of the transit and transit midpoint.\label{fig:calcium_planet}}
\end{center}
\end{figure*}

We investigate if the \calcium\ is an extended ionized atmosphere of the \target. We evaluate the behavior of the \calcium~IRT transmission signal in both the star and planet's rest frames (Figure~\ref{fig:calcium_planet}) and evaluate the transmission signal for \halpha\ (rightmost column). The variations in \calcium~IRT are seen to vary with planet orbital phase in both rest frames. We can rule-out that the \calcium~IRT is part of an extended atmosphere from \target\ because it is seen in emission and is not at rest in the planet's rest frame. The \halpha\ line changes also are not at rest in the planet's rest frame and highlight the $\pm$~\vsini\ change in absorption over the night. 

Close-in planets interact with their host star's magnetic field lines, which have been seen to manifest as changes in flux from the cores of chromospherically driven lines. Such changes have previously shown variations with the planet's orbital phase \citep{cauley19}. \cite{shkolnik08} showed that the equivalent width of \calciumhk\ varied consistently with the orbital phase of HD 179949\,b over 5~years of observations, which is indicative of a magnetically-driven star-planet interaction. They also found a strong correlation between \calciumhk\ and the mean emission from the chromospheric core of the \calcium~IRT. Similar studies have been conducted for HD\,189733\,b, where no clear evidence of star-planet interactions were seen via a similar analysis of \calciumhk\ \citep{fares10}. Following studies using 10~years of spectral observations of HD\,189733\,b found \calciumhk\ variations corresponding to transits, but with a leading phase offset of $\Delta \phi = 40-53^\circ$ \citep{lanza2012, cauley18_spi}. 

It is possible the \calcium~IRT variations in our observations are driven by star-planet interactions. Given the increased magnetic activity of young stars and the close proximity of young planets, these interactions may be expected. However, with limited orbital phase coverage it is difficult to say if this is the case. Multiple transits of \target\ are required to determine if the \calcium~IRT varies regularly with orbital phase. Simultaneous observations \calciumhk\ would also prove useful here.

\section{Interpretation \& Future Work}\label{sec:future}

\subsection{Interpretation}

In Section~\ref{subsec:clv}, we explored the possibility of our observed \halpha\ trend to be the result of CLV. We can confidently rule this out as being the source of the equivalent width variations due to the small effect we find ($\approx 10^{-5}$) compared to our observed signal (8~ppt). 

There appears to be slight indications of a transit in \halpha\ (Figure~\ref{fig:halpha}) with an underlying long term stellar activity trend. However, there is no visible Doppler tomographic signal. The change in excess \halpha\ absorption is also consistent with that seen for other young stars (Figure~\ref{fig:veloce}). We also find no evidence of \halpha\ changes in the planet's frame of reference (Figure~\ref{fig:calcium_planet}, rightmost column). We also show such \halpha\ variations can be explained by a stellar surface with 20\% coverage by starspots and faculae (Figure~\ref{fig:starry}). A 20\% inhomogeneity coverage, like that at the start of our toy model, is possiblle for young stars.

With no previous constraints on an extended atmosphere from \target\ or \vtau\,d, it is difficult to address how much this could be contaminating our OOT observations at the beginning of the night. However, since the \halpha\ excess absorption variations are seen in other young stars and they can be explained solely by the presence of stellar inhomogeneities, we speculate the \halpha\ is most likely stellar in nature.

We only find the Doppler tomographic signal in \calcium~IRT. If this is driven by star-planet interactions, then it would be expected to be seen in other lines that originate in the chromosphere. Since our observations do not contain any other such lines (e.g. \calciumhk), we cannot definitively say this is the case. It is still unclear as to why we do not see variability in other strong lines, specifically the \sodium\ doublet (Figure~\ref{fig:others}).

\subsection{Future Work}

While we speculate on several independent scenarios, the ground truth could very possibly be some combination of all presented interpretations. The observations presented here alone are insufficient to distinguish between them. Multiple transits observed with both \calciumhk\ and \calcium~IRT could help answer why the Doppler tomographic signal is only seen in this feature of our observations. If it is the result of star-planet interactions, there will be periodic equivalent width variations corresponding to the period of \target\ \citep[e.g. HD\,179949\,b;][]{shkolnik08}.

Since \halpha\ is known to contaminate transmission spectra \citep{rackham18} via the presence of stellar inhomogeneities and young stars are extreme cases of these surfaces \citep[e.g.][]{gully17}, a detection of an extended \halpha\ atmosphere is very challenging. Multiple observations of \target\ transiting over different stellar surface phases need to be observed to try and recover this signal. Additional OOT observations may be required to understand the general activity of \halpha\ for this active young star. If the $5\%$ variations of \halpha\ from our observations are stellar in nature, which we believe they are, then more OOT observations will help build a better spectral template to compare IT observations to.

\subsubsection{\helium\ at 1083.3~nm}

Starspot and AR contamination has a much less significant effect ($\approx 0.1$\%) for \helium\ at 1083.3~nm \citep{Oklopcic18}. Several detections of atmospheres have been made through observations of \helium\ \citep[e.g.][]{mansfield18, spake18, allart19, paragas21}. \cite{guilluy20} conclude that planets around active stars are good candidates for searching for \helium\ atmospheric signatures and signs of atmospheric escape. This would also be ideal for young planets, not still embedded in debris disks \citep{hirano20}, as stellar inhomogeneities are lower in contrast at longer wavelengths \citep{Strassmeier09}. Additionally, simultaneous observations of \target\ in \halpha\ and \helium\ will allow us to further understand if the trend in our observations originates from the planet or is dominated by underlying stellar activity.

Ground-based follow-up of \target\ and other transiting planets in this system using NASA Keck + NIRSPEC could allow for the first detection of an extended young atmosphere in \helium. If the greatest variations we will see are driven by chromospheric lines, then transits of \target\ in the UV may prove more fruitful. Observations using the Hubble Space Telescope + Wide-Field Camera 3 would allow for a full characterization of planetary atmospheres and the high-energy flux from the young active host star. Future James Webb Space Telescope observations of transits of \target\ as well as \vtau\,b and d would offer unprecedented detail of the composition as well as characterizing the extension of the young atmospheres. Observations of all three planets would allow for a detailed comparison of atmospheric composition and mass loss for planets of different radii in the same harsh stellar environment.

\subsubsection{\vtau\ and \tess}

Additionally, \tess\ will observe \vtau\ during its extended mission from September - November 2021. Simultaneous spectroscopy with this guaranteed  photometry would allow for a confident removal of stellar activity and first potential detection of an extended \halpha\ or \helium\ atmosphere. The resulting \tess\ light curves will show much lower variability amplitudes by virtue of spot contrast that would not be obtainable by ground-based photometry at the \tess\ wavelength coverage. Additionally, \tess\ overlaps in wavelength coverage \citep[600-1000~nm;][]{Ricker14} with \halpha, \calcium~IRT, and \helium, making it easier to compare the spectral response in simultaneous data sets.

X-ray observations of HD\,189733 provided evidence of an intense flaring event 3~ks after the eclipse of HD\,189733\,b \citep{wolk11}. These observations lead to speculation that the flare was the result of magnetic interactions between the star and planet. There is evidence of flares from \vtau\ in the \ktwo\ light curve \citep{david19_v1298all} and future \tess\ light curves at a higher cadence may display similar flare trends. Any phase dependence of flares with planet transits could be evidence of star-planet interactions and are unlikely dependent on stellar surface spot coverage \citep{feinstein20}.

\section{Conclusions} \label{sec:conclusion}

We have presented a Doppler tomographic and spectral analysis of \vtau, a 23~Myr K star, during a transit of its innermost known planet, \target.

\begin{enumerate} 

    \item We measure an alignment of \target\ of $\lambda =$~\obliquity\ from the \calcium~IRT at 849-866~nm. This is the only spectral feature with an obvious transit signal, hinting that measuring the spin-orbit alignment of young planets may be more feasible at near-infrared/infrared wavelengths. 
    
    \item \target\ adds to a growing list of young planets with low obliquities, including \vtau\,b \johnsoncitep, indicating these planets undergo smooth migration shortly after they formed. Multi-planet systems tend towards being well-aligned at all ages \citep{Albrecht13} and we have demonstrated that the \vtau\ system is no exception.
    
    \item Variations in \calcium~IRT could be an indication of star-planet interactions. More transit observations of \target\ in \calciumhk\ and \calcium~IRT may be able to firmly assess whether this is the case.
    
    \item The excess absorption in \halpha\ corresponds well to variations seen in other stars of younger and slightly older ages. The variations can also be replicated solely via the presence of starspots and faculae, leading us to believe stellar activity is the origin of the signal and it is not from an extended young atmosphere.
    
    \item These observations alone are insufficient to characterize any potential extended atmosphere from \target. More transits are required to disentangle stellar from planetary signals. Future observations in the UV, for more chromospheric lines, and NIR/IR, where starspots are at lower contrast, may prove more optimistic for characterizing young planetary systems. 
    
    \item There is still much work to be done to understand the nature of the variability seen in our spectra. The data\footnote{Gemini Program GN-2019B-FT-215 on the Gemini Data Archive: \url{https://archive.gemini.edu/searchform/}} used in this analysis can be found on the Gemini Archive. In the interests of open science, our analysis code is also publicly available\footnote{\url{https://github.com/afeinstein20/doppler_tomography}} and can be found via the \faCloudDownload\ icons throughout the paper.

\end{enumerate}

\acknowledgements

We thank Fred Ciesla and Darryl Seligman for thoughtful conversations which improved the presented results. We thank the anonymous reviewer for improving the quality and clarity of this work. This work is based on observations obtained through the Gemini Remote Access to CFHT ESPaDOnS Spectrograph (GRACES). ESPaDOnS is located at the Canada-France-Hawaii Telescope (CFHT), which is operated by the National Research Council of Canada, the Institut National des Sciences de l’Univers of the Centre National de la Recherche Scientifique of France, and the University of Hawai’i. ESPaDOnS is a collaborative project funded by France (CNRS, MENESR, OMP, LATT), Canada (NSERC), CFHT and ESA. ESPaDOnS was remotely controlled from the international Gemini Observatory, a program of NSF’s NOIRLab, which is managed by the Association of Universities for Research in Astronomy (AURA) under a cooperative agreement with the National Science Foundation on behalf of the Gemini partnership: the National Science Foundation (United States), the National Research Council (Canada), Agencia Nacional de Investigación y Desarrollo (Chile), Ministerio de Ciencia, Tecnología e Innovación (Argentina), Ministério da Ciência, Tecnologia, Inovações e Comunicações (Brazil), and Korea Astronomy and Space Science Institute (Republic of Korea). Data were obtained under the program GN-2019B-FT-215. 

This work was enabled by observations made from the Gemini North telescope, located within the Maunakea Science Reserve and adjacent to the summit of Maunakea. We are grateful for the privilege of observing the Universe from a place that is unique in both its astronomical quality and its cultural significance.

Based in part on data acquired at the Anglo-Australian Telescope under program A/2020B/09. We acknowledge the traditional owners of the land on which the AAT stands, the Gamilaroi people, and pay our respects to elders past and present.

ADF acknowledges support from the National Science Foundation Graduate Research Fellowship Program under Grant No. (DGE-1746045). Any opinions, findings, and conclusions or recommendations expressed in this material are those of the author(s) and do not necessarily reflect the views of the National Science Foundation. 

This research has made use of NASA's Astrophysics Data System Bibliographic Services.

\software{%
    numpy \citep{numpy},
    matplotlib \citep{matplotlib},
    scipy \citep{jones01},
    astropy \citep{astropy:2013, astropy18},
    ccdproc \citep{ccdproc},
    OPERA pipeline \citep{martioli12, teeple14},
    starry \citep{Luger19},
    batman \citep{kreidberg15},
    \stella\ \citep{feinstein20_joss},
    \banyan\ \citep{gagne18},
    \texttt{MISTTBORN} \citep{johnson17, johnson18},
    emcee \citep{Foreman-Mackey13}
    }

\facility{
    Gemini-North: GRACES \citep{chene14},
    Anglo-Australian Telescope: Veloce-Rosso Spectrograph \citep{veloce_rosso}
}

\bibliography{exopapers}

\end{document}